\documentclass[ALICE,manyauthors]{cernphprep}

\usepackage{hyperref}
\usepackage{graphicx}  
\usepackage{dcolumn}   
\usepackage{bm}        
\usepackage{amssymb}   
\usepackage{amsfonts}
\usepackage{graphics}
\usepackage{grffile}   
\usepackage{epsfig}
\usepackage{units}
\usepackage[usenames]{color}
\usepackage[normalem]{ulem} 
\usepackage{color}
\usepackage[utf8]{inputenc}
\usepackage[T1]{fontenc}
\usepackage{subfigure}
\usepackage{lineno}
\usepackage{cite} 
\usepackage{orcidlink}

\usepackage{enumitem}
\setlist{itemsep=0pt, topsep=1.5mm, leftmargin=8mm}

\newcommand{\pp}{\ensuremath{\mathrm {p\kern-0.05em p}}}
\newcommand{\PbPb}{\ensuremath{\mbox{Pb--Pb}}}
\newcommand{\GeVc}{\ensuremath{\mathrm{GeV}\kern-0.05em/\kern-0.02em c}}
\newcommand{\sqrts}{\ensuremath{\sqrt{s_{\mathrm{NN}}}}}
\newcommand{\pT}{\ensuremath{p_{\mathrm{T}}}}
\newcommand{\kT}{\ensuremath{k_{\mathrm{T}}}}

\newcommand{\pTchjet}{\ensuremath{p_{\mathrm{T}}^{\rm{ch\; jet}}}}
\newcommand{\pTtruth}{\ensuremath{p_{\mathrm{T,truth}}^{\mathrm{ch\; jet}}}}
\newcommand{\pTdet}{\ensuremath{p_{\mathrm{T,det}}^{\mathrm{ch\; jet}}}}
\newcommand{\Rmax}{\ensuremath{R_{\mathrm{max}}}}
\newcommand{\pTsubjet}{\ensuremath{p_{\mathrm{T}}^{\rm{ch\; subjet}}}}
\newcommand{\zr}{\ensuremath{z_r}}
\newcommand{\zrNP}{\ensuremath{z_r^{\mathrm{NP}}}}
\newcommand{\zrtruth}{\ensuremath{z_{r\mathrm{, truth}}}}
\newcommand{\zrdet}{\ensuremath{z_{r\mathrm{, det}}}}

\newcommand{\etajet}{\ensuremath{\eta_{\mathrm{jet}}}}
\newcommand{\Ninc}{\ensuremath{N_{\mathrm{jet}}}}
\newcommand{\Sinc}{\ensuremath{\sigma_{\mathrm{jet}}}}

\begin{document}


\begin{titlepage}
\PHyear{2022}
\PHnumber{060}      
\PHdate{22 March}  
%

\title{Measurement of inclusive and leading subjet fragmentation in pp and $\mbox{Pb--Pb}$ collisions at $\bf{\sqrt{\textit{s}_{NN}}=5.02}$ TeV}
\ShortTitle{Measurement of inclusive and leading subjet fragmentation}   

\Collaboration{ALICE Collaboration\thanks{See Appendix~\ref{app:collab} for the list of collaboration members}}
\ShortAuthor{ALICE Collaboration} 

\date{\today}
\begin{abstract}

This article presents new measurements of the fragmentation properties of jets 
in both proton--proton (pp) and heavy-ion collisions with the ALICE experiment at the Large Hadron Collider (LHC). 
We report distributions of the fraction $z_r$ of transverse momentum 
carried by subjets of radius $r$ within jets of radius $R$.
Charged-particle jets are reconstructed at midrapidity using the anti-$k_{\rm{T}}$ algorithm with jet radius $R=0.4$, and subjets are reconstructed by reclustering the jet constituents 
using the anti-$k_{\rm{T}}$ algorithm with radii $r=0.1$ and $r=0.2$.
In proton--proton collisions, we measure both the inclusive and leading subjet distributions.
We compare these measurements to perturbative calculations at next-to-leading logarithmic accuracy, which suggest a large impact of threshold resummation and hadronization effects on the $z_r$ distribution.
In heavy-ion collisions, 
we measure the leading subjet distributions, which allow access to a region of harder jet fragmentation than has been probed by previous measurements of jet quenching via hadron fragmentation distributions.
The $z_r$ distributions enable extraction of the parton-to-subjet fragmentation function
and allow for tests of the universality of jet fragmentation functions in the quark--gluon plasma (QGP).
We find no significant modification of $z_r$ distributions in $\mbox{Pb--Pb}$ compared to $\mathrm{p\kern-0.05em p}$ collisions.
However, the distributions are also consistent with a hardening trend for $z_r<0.95$, as predicted by several jet quenching models. 
As $z_r \rightarrow 1$ our results indicate that any such hardening effects
cease, exposing qualitatively new possibilities to disentangle competing jet quenching mechanisms.
By comparing our results to theoretical calculations based on an independent extraction of the
parton-to-jet fragmentation function, we find consistency with the universality of jet fragmentation
and no indication of factorization breaking in the QGP. 

\end{abstract}

\end{titlepage}
\setcounter{page}{2}


\section{Introduction}

Measurements of high-energy jets produced in scattering experiments 
offer opportunities to test perturbative calculations in quantum chromodynamics (QCD)~\cite{Larkoski:2017jix,Asquith:2018igt,Marzani:2019hun, ATLAS:2019mgf, CMS:2018ypj, ALICE:2021njq, STAR:2020ejj, Chen:2021uws, H1:2021wkz}
and to probe the properties of the quark--gluon plasma (QGP)~\cite{Bjorken:1982qr, STAR:2005gfr, PHENIX:2004vcz, Muller:2012zq, Braun-Munzinger:2015hba, Busza:2018rrf}.
Heavy-ion collisions can be used to produce short-lived droplets of QGP, 
serving as a laboratory system to study the emergence
of this high-temperature, strongly-coupled, deconfined system of quarks and gluons in QCD.
While several of its transport coefficients have been constrained by experimental measurements~\cite{Bernhard:2019bmu, Parkkila:2021tqq, Nijs:2020ors, JETSCAPE:2020mzn, Liu:2016ysz, Scardina:2017ipo, Xu:2017obm, JET:2013cls, Andres:2016iys, Xie:2019oxg, JETSCAPE:2021ehl}, the detailed physical properties of the QGP, including the nature of its degrees of freedom as a function of resolution scale, remain unknown.

Properties of the QGP can be inferred by comparing jets in heavy-ion collisions, which traverse the QGP, to their counterparts in proton--proton collisions. Significant experimental and theoretical effort has been made to measure and calculate the modification of jet observables in heavy-ion collisions, known as jet quenching, in an ongoing attempt to achieve a unified description of the jet--QGP interaction~\cite{Qin:2015srf, Blaizot:2015lma, Majumder:2010qh}.
It has been established that jets traversing the QGP emit soft medium-induced radiation outside of the jet cone,
causing their observed yields to be significantly suppressed as compared to \pp{} collisions,
and that the fragmentation pattern of the resulting observed jets can be both narrowed and hardened in certain regions of phase space~\cite{Acharya:2019jyg, ATLAS:2018gwx, CMS:2021vui, STAR:2020xiv, ALICE:2021obz, ALICE:2018dxf, ATLAS:2018bvp, ATLAS:2019dsv, CMS:2014jjt, CMS:2018mqn, CMS:2017qlm, CMS:2018fof, ALICE:2015mdb, STAR:2017hhs}.
However, the precise role of several theoretical mechanisms of jet quenching, such as
quark vs. gluon energy loss, factorization breaking, and color coherence remains unclear. 
It is essential to understand and disentangle these mechanisms in order to use jet quenching
to reveal physical properties of the QGP.

\begin{figure}[!t]
\centering{}
\includegraphics[scale=0.34]{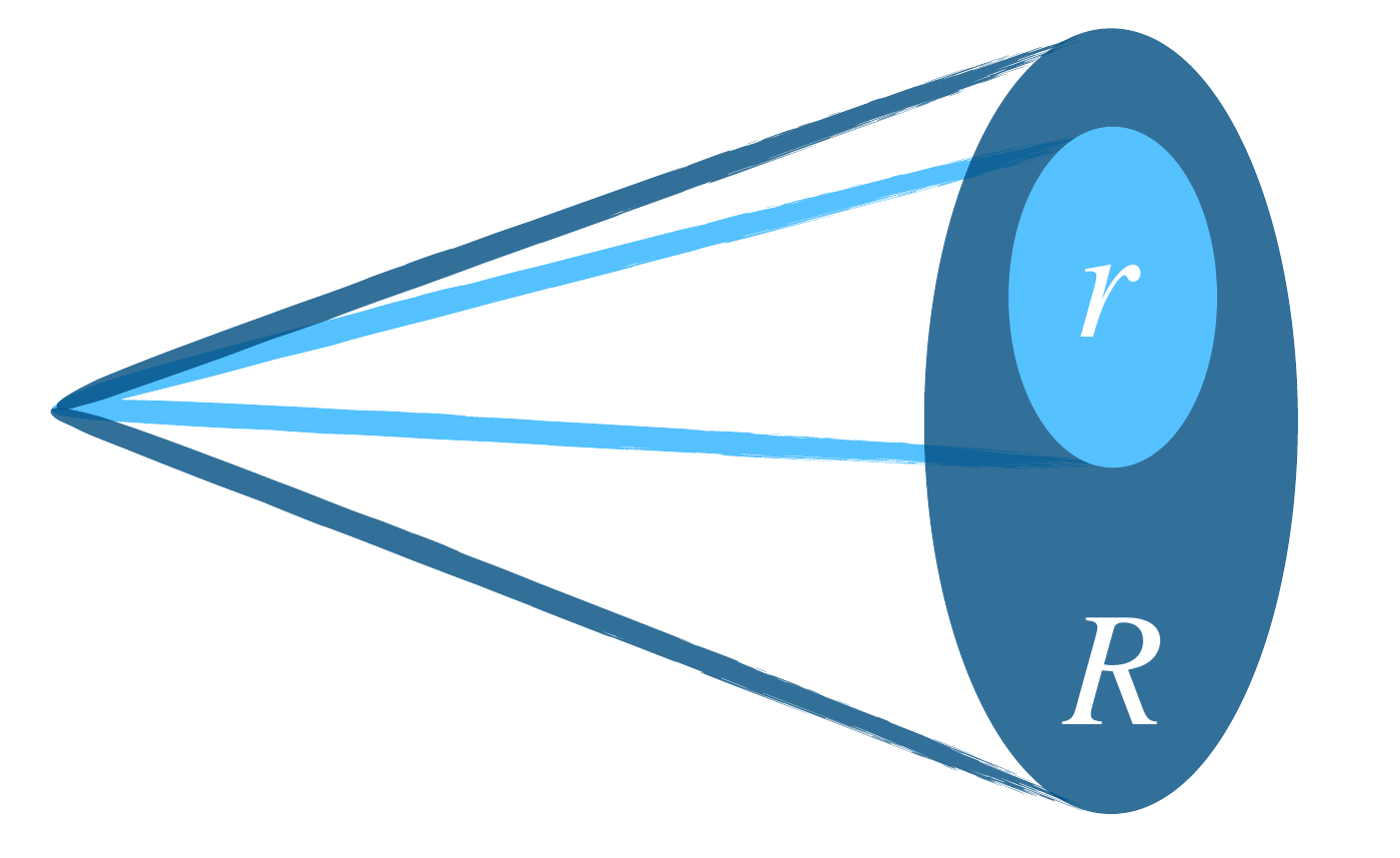}
\caption{Cartoon of a subjet of radius $r$ inside a jet of radius $R$. We consider charged-particle subjets clustered using the anti-\kT{} algorithm from the constituents of inclusive charged-particle jets.}
\label{fig:cartoon}
\end{figure}

In this article, we consider measurements of charged-particle \textit{subjets},
defined by first clustering inclusive charged-particle jets with the 
anti-\kT{} algorithm~\cite{Cacciari:2008gp} with radius $R$,
and then reclustering the jet constituents with the anti-\kT{} algorithm with subjet radius $r<R$, as illustrated in Fig.~\ref{fig:cartoon}~\cite{Dai:2016hzf}.
We focus on the fraction of charged-particle transverse momentum (\pT) carried by the subjet:
\[
\zr = \frac{\pTsubjet}{\pTchjet},
\]
where ${p_{\mathrm{T}}^{\rm{ch\; (sub)jet}}}$ is the transverse momentum of the charged-particle (sub)jet.

In \pp{} collisions, both the inclusive and leading subjet \zr{} distributions have been calculated 
perturbatively for a variety of $r$ and $R$ values~\cite{Kang:2017mda, Neill:2021std}.
These calculations involve several interesting aspects that can be studied experimentally such as
the role of threshold resummation, and, in the leading subjet case,
nonlinear evolution of the jet fragmentation function.
Similar effects have recently been examined in e$^+$e$^-$ collisions~\cite{Chen:2021uws, Neill:2021std}.

In heavy-ion collisions, subjet observables have been proposed as
sensitive probes of jet quenching~\cite{Kang:2017mda, Neill:2021std, Apolinario:2017qay, Caucal:2020xad}.
The subjet \zr{} observable presents several unique opportunities to study jet quenching:
\begin{enumerate}[itemsep=-0.5ex,topsep=4pt]

    \item \textit{Probe high-$z$ fragmentation}.
    The subjet fragmentation distribution \zr{} is complementary to the longitudinal
    momentum fraction $z$ of hadrons in jets~\cite{ATLAS:2018bvp, ATLAS:2019dsv, CMS:2014jjt, CMS:2018mqn}. The subjet fragmentation distribution can be understood as a generalization of the hadron fragmentation distribution, where in the limit $r\rightarrow0$ the two become equal. 
    By generalizing to $r>0$, subjet measurements offer the benefit of probing higher $z$ values than hadron measurements; even for small subjet aperture the \zr{} distribution exhibits a peak at large \zr{}, whereas the hadron $z$ distribution populates much lower $z$ values.
    This enables a more precise handle on the overall hardness of jet fragmentation,
    and thereby makes it possible to access a quark-dominated sample of jets~\cite{Neill:2021std} at high \zr{}. This in turn provides
    an opportunity to expose the interplay of soft medium-induced radiation with the relative suppression of gluon vs. quark jets, and introduces a method to do so based on inclusive jet samples alone, complementary to existing methods comparing inclusive and photon-tagged jet observables~\cite{ATLAS:2019dsv, CMS:2018mqn}.

    \item \textit{Test the universality of jet fragmentation in the QGP}. 
    In vacuum, it is expected that the parton-to-jet fragmentation function, $J(z)$,
    is equal to the parton-to-subjet fragmentation function $J_{r}(z)$~\cite{Kang:2017mda}.
    However, it is unknown whether this universality of jet fragmentation functions holds in the QGP independently of the observable considered~\cite{Qiu:2019sfj}.
    Measurements of \zr{} distributions are directly sensitive to the medium-modified parton-to-subjet fragmentation function, $J_{r,\rm{med}}(z)$, and can be used to extract it. 
    The extracted $J_{r,\rm{med}}(z)$ can then be compared
    to an independently extracted medium-modified parton-to-jet fragmentation function, $J_{\rm{med}}(z)$, 
    to test the universality of in-medium jet fragmentation across different jet observables in heavy-ion collisions and look for signs of factorization breaking.

    \item \textit{Measure energy loss at the cross section level}.
    A well-defined method of measuring out-of-cone energy loss at the cross section level has been proposed by computing moments of the leading subjet \zr{} distribution~\cite{Neill:2021std}.
    The first moment or ``subjet energy loss'' describes the fraction of jet \pT{} not carried by the leading subjet, and higher moments describe fluctuations in this energy loss. 
    These quantities can be computed in both \pp{} and \PbPb{} collisions for a variety of $r$ and $R$ values,
    and contrasted with other measures of jet modification.
    
\end{enumerate}

\section{Experimental setup and data sets}\label{sec:setup}
A description of the ALICE detector and its performance can be found in
Refs.~\cite{aliceDetector, ALICE:2014sbx}. The data sample of \pp{} collisions used in this
analysis was collected in 2017 during the LHC Run 2 at $\sqrt{s} = 5.02$ TeV 
using a minimum-bias (MB) trigger defined by the coincidence of the signals from the two
V0 scintillator arrays in the forward region~\cite{ALICE:2013axi}. 
The \PbPb{} data set was collected in 2018 at $\sqrts=5.02$ TeV. 
A central collision trigger was used which selects events in the 0--10\% centrality interval based on the
multiplicity of produced particles in the V0 detector acceptance~\cite{ALICE:2013hur, centrality502}.
In the event, the primary vertex was required to be within 10 cm along the beam axis from the center of the detector. 
Beam-induced background events were removed using timing information from the V0 detectors
and, in \PbPb{} collisions, from two neutron
Zero Degree Calorimeters located $\pm112.5$ m along the beam axis 
from the center of the detector.
Pileup events were rejected based on multiple reconstructed vertices and 
tracking selections~\cite{Acharya:2019jyg}.
After these selections, the pp data sample
contains 870 million events and corresponds to an integrated luminosity of
$18.0 \pm 0.4$ nb$^{-1}$~\cite{ppXsec}. The \PbPb{} data sample contains 92 million events in 0--10\%
central collisions, corresponding to an integrated luminosity of approximately 0.12 nb$^{-1}$. 

This analysis uses charged-particle tracks reconstructed based on the information from the Time Projection Chamber
(TPC)~\cite{Alme_2010} and the Inner Tracking System (ITS)~\cite{Aamodt:2010aa}.
While track-based observables are collinear-unsafe~\cite{Chang:2013rca, Chen:2020vvp, Chien:2020hzh},
they can be measured with greater precision than calorimeter-based observables,
and recent measurements have demonstrated that for many substructure observables 
track-based distributions are compatible with the corresponding collinear-safe distributions~\cite{ATLAS:2019mgf}.
We define two types of tracks: global tracks and complementary tracks~\cite{ALICE:2014sbx}.
Global tracks are required to include at least
one hit in the Silicon Pixel Detector (SPD) comprising the
two innermost layers of the ITS and to satisfy multiple criteria on the quality of track reconstruction in the TPC and its pointing to the collision vertex.
Complementary tracks are all those satisfying all the selection
criteria of global tracks except for the request of a point in the SPD. They are refitted using
the primary vertex to constrain their trajectory in order to preserve good momentum
resolution, especially at high transverse momentum (\pT{}).
Including this second class of tracks ensures approximately uniform azimuthal acceptance, 
while providing similar \pT{} resolution to tracks with SPD hits. 
Tracks with $\pT>0.15\;\GeVc$ were 
accepted over the pseudorapidity range $|\eta| < 0.9$ and azimuthal angle $0 < \varphi < 2\pi$. 
The momentum resolution $\sigma(\pT)/\pT$ of the accepted tracks was estimated from the covariance matrix of the track fit parameters~\cite{ALICE:2014sbx},
and is approximately 1\% at track $\pT=1$ \GeVc{} and 4\% at ${\pT=50 \;\GeVc}$.

The instrumental performance of the detector was estimated 
with a simulation performed using PYTHIA8 Monash 2013~\cite{Sjostrand:2014zea, Skands:2014pea}
 for the event generation and the GEANT3 transport code~\cite{Brun:1119728} to propagate particles through the simulated ALICE apparatus.
The tracking efficiency in \pp{} collisions is approximately 67\% 
at track $\pT=0.15 \;\GeVc$, and rises to approximately 84\% at $\pT=1 \;\GeVc$, and remains above 75\% at higher \pT.
Studies of the centrality dependence of the tracking efficiency in a HIJING~\cite{Gyulassy:1994ew} simulation demonstrated that the tracking efficiency 
is approximately 2\% lower in 0--10\% central \PbPb{} collisions compared to \pp{} collisions, independent of track \pT.

\section{Analysis method}\label{sec:analysis}
Jets were reconstructed from charged-particle tracks with FastJet 3.2.1~\cite{Cacciari:2011ma} using the
anti-$k_{\mathrm{T}}$ algorithm with $E$-scheme recombination
and radius (or resolution parameter) $R=0.4$~\cite{Cacciari:2008gp, Cacciari:2008gn}.
Subjets were reconstructed by reclustering the jet constituents using the 
anti-$k_{\mathrm{T}}$ algorithm with $E$-scheme recombination
and radii $r=0.1$ and $r=0.2$.
The pion mass was assumed for all jet constituents.
Jets containing tracks with $\pT>100\;\GeVc$, 
corresponding to $<1\%$ of the jet sample in the considered kinematic range,
were discarded in order to ensure good momentum determination.
Jets in heavy-ion collisions have a large uncorrelated background contribution due
to the underlying event (UE)~\cite{ALICE:2012nbx}.
The event-by-event constituent subtraction method was used before jet finding was performed~\cite{Berta:2014eza, Berta:2019hnj}.
This corrects the overall jet \pT{} and its substructure simultaneously by subtracting
UE energy constituent by constituent. 
A maximum recombination distance $\Rmax = 0.25$ was used~\cite{Berta:2019hnj}.
In \pp{} collisions, the UE was not subtracted and must be included as a model component in all theory comparisons.
According to PYTHIA8, the impact of the UE in \pp{} collisions on the leading \zr{} distribution is <3\% for $\zr<0.95$, and grows up to 13\% as $\zr \rightarrow 1$.
The jet axis is required to be within the fiducial volume of the TPC, $\left| \etajet \right| < 0.9 - R$, where $\etajet$ is the jet axis pseudorapidity.

The jet reconstruction performance is studied by
simulating \pp{} events and particle transport through the ALICE detector material
as described in Sec.~\ref{sec:setup}.
We compare PYTHIA8 generated jets at ``truth level'' (before the particles undergo interactions 
with the detector) to those at ``detector level'' (after detector simulation).
The truth-level jet was constructed from the charged primary
particles of the PYTHIA8 event, defined as all particles with a 
mean proper lifetime larger than 1 cm/$c$, and excluding the decay
products of these particles~\cite{primaryParticleALICE}.
For the \PbPb{} data analysis, we embedded the simulated \pp{} events after track reconstruction into 0--10\% centrality \PbPb{} measured events to
account for background effects, and applied the constituent subtraction procedure described above.
A jet matching procedure is used in order to associate jets at the truth level 
to jets at the detector level.
In \pp{} collisions, this matching procedure is based on geometrically matching jets within $\Delta R < 0.6\;R$,
where $\Delta R = \sqrt{\Delta y ^2 + \Delta \varphi ^2}$ is their rapidity-azimuth separation.
In \PbPb{} collisions, the matching procedure includes both this geometrical requirement and 
further requires that the jet contains at least 50\% of the total track \pT{} of the associated reconstructed jet from the embedded \pp{} event at detector level.
To study the subjet \zr{} reconstruction performance, a similar matching procedure was adopted.
In order to match inclusive subjets at the truth level with their counterparts at the detector level,
we apply the same matching procedure as described above for jets, except with the subjet radius $r$
replacing the jet radius $R$ in the geometrical matching criteria.
In the case of leading subjets, however, we match the leading subjet at the truth level to the leading
subjet at the detector level, without requiring geometrical criteria.
The jet energy scale shift $(\pTdet - \pTtruth)/\pTtruth$ is a long-tailed asymmetric distribution due to tracking inefficiency with a peak at zero, corresponding to $\pTdet=\pTtruth$,
where \pTdet{} is the detector-level \pTchjet{}, and \pTtruth{} is the truth-level \pTchjet{}~\cite{ALICE:2013dpt}.
The mean jet energy scale shift for $R=0.4$ charged-particle jets
in \pp{} collisions is approximately $-13\%$ at $\pTtruth=20\;\GeVc$
and decreases to $-21\%$ at $\pTtruth=100\;\GeVc$.
The jet energy resolution is approximately constant at $21\%$ in \pp{} collisions; in \PbPb{} collisions an additional contribution from
the underlying event fluctuations further broadens the distribution of the 
jet energy scale shift, with the standard deviation of the 
background fluctuation contribution to the shift
$\sigma_{\delta \pT{}}$~\cite{ALICE:2012nbx} being equal to approximately $11\;\GeVc$, independent of \pTtruth{}.
The \zr{} reconstruction performance behaves qualitatively similarly as the jet \pT{} 
described above, with a peak near zero in the relative residual distribution corresponding to $\zrdet=\zrtruth$.
The \zr{} reconstruction resolution is $\mathcal{O}(10\%)$, with 
asymmetric tails that are broader for small \zr{} than for large \zr{}, 
and broader in \PbPb{} compared to pp{} collisions.

Local fluctuations in the underlying event of a heavy-ion collision can result in 
an incorrect subjet (unrelated to the hard scattering) being 
identified by the reclustering algorithm.
This ``mistagging'' effect is in exact analogy to the 
case of identifying groomed jet splittings in the presence of a large underlying event~\cite{Mulligan:2020tim},
although with improved robustness to mistagging effects~\cite{STAR:2021kjt}.
In order to address this issue, the measurement was
performed by restricting to $\zrdet>0.5$, which mitigates these effects.
The subjet purity due to these background effects was evaluated by embedding 
jets simulated with the PYTHIA8 event generator~\cite{Sjostrand:2014zea} into measured \PbPb{} collisions and following the procedure in Ref.~\cite{Mulligan:2020tim} .
The residual background
contribution remains below 5\% for $\zr>0.6$ and increases up to approximately 20\% at $\zr=0.5$,
for the \pTdet{} range considered in this measurement.
This level of background contamination is small enough to allow
the results to be unfolded for detector effects and background fluctuations.
The corresponding contamination in \pp{} collisions is negligible.

The reconstructed \pTchjet{} and \zr{} differ from
their true values due to tracking inefficiency,
particle-material interactions, and track \pT{} resolution. Moreover,
in \PbPb{} collisions, background fluctuations significantly smear
the reconstructed distributions of  \zr{}. To account for
these effects, we simulated events at the truth level and detector level
and matched both jets and subjets at the truth level and detector level
as described above. 
We constructed a 4D response matrix (RM) that describes
the detector and background response in \pTchjet{} and \zr{}: 
$R\left( \pTdet, \pTtruth, \zrdet, \zrtruth \right)$.
For both the \pp{} and \PbPb{} results, we use PYTHIA8 to generate the RM since
previous studies indicate that the effect of changing fragmentation patterns on the response is small. This is due to the fact that there is only an indirect dependence via the response matrix~\cite{CMS:2014jjt, ATLAS:2017xfa, ALICE:2021obz}; an uncertainty due to this model dependence is assessed in Sec.~\ref{sec:sys}.
A simultaneous unfolding was then performed in \pTchjet{} and \zr{} using the iterative Bayesian
unfolding algorithm~\cite{dagostini2010improved, DAgostini} implemented in the RooUnfold package~\cite{roounfold}.
The prior distribution for the unfolding procedure is taken to be the truth-level distribution from PYTHIA8.
In \PbPb{} collisions, lower limits of $\pTdet > 60\;\GeVc$ and $\zrdet>0.5$ are imposed 
on the data that is input to the unfolding, in order to reject combinatorial jets and mistagged subjets.
No such limitation is imposed on \pTtruth{} or \zrtruth{} during the unfolding process.
The distributions were corrected for ``misses'', in which a jet was generated inside the considered truth-level range but not inside the detector-level range.
In \pp{} collisions the rate of misses is $<3\%$, whereas in \PbPb{} collisions the rate of misses ranges from $24-40\%$ due to the aforementioned fluctuations in the UE.
The rate of ``fakes'', in which a jet exists inside the considered detector-level range but not inside the truth-level range,
is $<1\%$ and therefore negligible.
The number of iterations, which sets the strength of regularization, was
chosen by minimizing the quadratic
sum of the statistical and systematic unfolding uncertainties described in Sec.~\ref{sec:sys}.
This results in the optimal number of iterations equal to 3 in all cases. 

To validate the performance of the unfolding procedure, we performed
refolding tests, in which the RM is multiplied by the unfolded solution
and compared to the original detector-level spectrum. We also did closure tests,
in which the shape of the input MC spectrum
is modified to account for the possibility that the true distribution may
be different from the MC input spectrum (using the same scalings as for the systematic variations in the unfolding prior described below in Sec.~\ref{sec:sys}). 
In all cases, successful closure was obtained within statistical uncertainties.
Additionally, we performed a closure test to quantify the 
sensitivity of the final result to combinatorial jets and background
subjets. This consisted of redoing the entire analysis on “combined” events containing a PYTHIA8
event and a thermal background, in which “combined” jets were
clustered from the combination of PYTHIA8 detector-level particles and thermal background particles.
The background was modeled by generating $N$ particles with \pT{} taken from a Gamma distribution,
$f_\Gamma \left( \pT;\beta \right) \propto \pT e^{-\pT/\beta}$,
where $N$ and $\beta$ were fixed to roughly fit the $R=0.4$ $\delta \pT{}$ distribution in 
\PbPb{} data~\cite{ALICE:2012nbx}.
This background model was verified to describe the subjet purity to percent-level accuracy.
The test consisted of constructing the combined detector-level jet spectrum, building the
RM, unfolding the combined jets, and comparing the spectrum to the truth-level PYTHIA8 spectrum. 
Because the background does not contain hard jets, this test is able to quantify the
extent to which the analysis procedure recovers the signal distribution and is free of background contamination.
The unfolded combined jet spectrum was found to be consistent with the truth-level spectrum within statistical uncertainties, thus confirming a successful closure.

\section{Systematic uncertainties}\label{sec:sys}

The systematic uncertainties in this measurement are due to the tracking
efficiency, the unfolding procedure, the model dependence of the event generator, and in the case of \PbPb{} collisions, 
the background subtraction procedure.
Table~\ref{tab:sysunc} summarizes the systematic uncertainty contributions 
for \pp{} and \PbPb{} collisions for the two considered values of $r$.
The total systematic uncertainty is calculated as the sum in quadrature of all of the individual sources described below.

The systematic uncertainty due to the uncertainty in tracking efficiency is evaluated by randomly discarding charged tracks before jet finding. 
The tracking efficiency uncertainty, estimated from the variation of the track selection criteria and a detailed study of the ITS--TPC track-matching efficiency uncertainty, is 4\%.
In order to assign a systematic uncertainty to the final
result, we constructed an alternative RM with this random track rejection
and repeated the unfolding procedure. 
The result was compared to the default result, with
the differences in each bin taken as the systematic uncertainty. 
The uncertainty on the track momentum resolution is a sub-leading
effect to the tracking efficiency and is negligible.

Several variations of the unfolding procedure are performed
in order to estimate the systematic
uncertainty arising from the unfolding correction:
\begin{itemize}
    \item The number of iterations in the
    unfolding was varied by $\pm2$ units and the average difference
    with respect to the nominal result is taken as the systematic uncertainty.
    \item The prior distribution was simultaneously scaled by $(\pTchjet)^{\pm0.5}$
     and a linear scaling in \zr{} by $\pm 50\%$ over its reported range.  
    The average difference between the result unfolded with this prior and the original 
    is taken as the systematic uncertainty. 
    \item The detector-level binnings in \zr{} were varied to be finer and coarser than the nominal binning.
    \item The lower bound in the detector level charged-particle jet transverse momentum \pTdet{} range was varied by $\pm5$ \GeVc. 
\end{itemize}
The total unfolding systematic uncertainty is then the standard deviation of the results from the variations, $\sqrt{\sum_{i=1}^{N} \sigma_{i}^{2} / N}$, 
where $N=4$ and $\sigma_{i}$ is the systematic uncertainty due to a single variation, since they each comprise independent measurements of the same underlying systematic uncertainty in the unfolding correction.

\begin{table}[!b]
\centering
\caption{Summary of systematic uncertainties on unfolded \zr{} distributions for $80<\pTchjet<120\;\GeVc$.
The ranges correspond to the minimum and maximum systematic uncertainties obtained.}
\begin{tabular}{ l cccccc }
\tabularnewline \hline \hline & \multicolumn{5}{c}{Relative uncertainty (\%)}
\tabularnewline \hline \pp{} & Trk. eff. & Unfolding & Generator &  & Total 
\\
$r = 0.1$ & 0--8\% & 1--6\% & 0--4\% && 2--10\% \\
$r = 0.2$ & 0--10\% & 1--7\% & 0--3\% && 1--11\% \\
\hline \PbPb{} 0--10\% & Trk. eff. & Unfolding & Generator & Bkgd. sub. & Total 
\\
$r = 0.1$ & 1--24\% & 3--17\% & 1--22\% & 5--10\% & 10--33\% \\
$r = 0.2$ & 1--18\% & 1--10\% & 1--20\% & 1--6\% & 7--25\% \\
\hline \hline

\end{tabular} 
\label{tab:sysunc}
\end{table}

The constituent subtraction introduces a bias in the observed distributions,
since it implicitly makes a choice of how much \pT{} to subtract from soft particles compared to the hard particles, and similarly for their angular distributions.
To estimate the size of the systematic uncertainty related to the background subtraction, 
we varied \Rmax{} from ``under-subtraction'' ($\Rmax=0.05$)
to ``over-subtraction'' ($\Rmax=0.7$), around the nominal value of $\Rmax=0.25$. 
The maximum deviation of these two variations was assigned as the systematic uncertainty.

The systematic uncertainty due to the model-dependence of the generator used to construct the 
RM is estimated by comparing results obtained with PYTHIA8~\cite{Sjostrand:2014zea}
to those obtained with  HERWIG7~\cite{Bellm:2015jjp} (in the \pp{} case)
or JEWEL 2.2.0~\cite{Zapp:2012ak, Zapp:2013vla} (in the \PbPb{} case).
For HERWIG7, the default tune was used, and for JEWEL, we adopted the settings described in 
Ref.~\cite{KunnawalkamElayavalli:2017hxo}, with an initial temperature $T_{\rm{i}}=590\;\rm{MeV}$ and no recoils.
These RMs are then used to unfold the measured
data, and the differences between PYTHIA8 and HERWIG7 (in the \pp{} case) or PYTHIA8 and JEWEL (in the \PbPb{} case) are taken as a symmetric uncertainty.
We note that in the \PbPb{} case, the results continue to exhibit little-to-no-modification when unfolded with the JEWEL-based response matrix despite that JEWEL itself exhibits large modifications as shown in Figs.~\ref{fig:zr-AA-r01},\ref{fig:zr-AA-r02} confirming that the experimental results remain under good control despite the large variations in the jet quenching models themselves.

\section{Results}
We report the \zr{} distributions for $r=0.1$ and $r=0.2$ in 
both \pp{} and \PbPb{} collisions.
All presented results use $R=0.4$ jets reconstructed from charged particles at midrapidity,
and are corrected for detector effects and (in \PbPb{} collisions) underlying-event fluctuations.
We report results for \pTchjet{} between 80 and 120 GeV/$c$ in both \pp{} and \PbPb{} collisions,
as well as a result with finer binning in \zr{} for \pTchjet{} between 100 and 150 GeV/$c$ in \PbPb{} collisions.
The distributions are reported as normalized differential cross sections,
\begin{equation} \label{eq:1}
\frac{1}{\Sinc} \frac{\mathrm{d}\sigma}{\mathrm{d}\zr}
=
\frac{1}{\Ninc} \frac{\mathrm{d}N} {\mathrm{d}\zr},
\end{equation}
where \Ninc{} (\Sinc{}) is the number (cross section) of inclusive charged-particle jets within the given \pTchjet{} interval, and $N$ ($\sigma$) is the number (cross section) of subjets.
With this normalization, the integral of Eq. (\ref{eq:1})
is equal to the average number of subjets per jet:
\begin{equation} \label{eq:2}
\left<N^{\mathrm{subjets}}\right> = \int_0^1  \mathrm{d}\zr \frac{1}{\Sinc} \frac{\mathrm{d}\sigma}{\mathrm{d}\zr}.
\end{equation}

Using the leading subjet distributions, we also compute the average ``subjet energy loss'':
\begin{equation} \label{eq:3}
\langle z_{\rm{loss}}\rangle =1-\int _0^1\mathrm{d}z_r\;z_r \frac{1}{\sigma}\frac{\mathrm{d}\sigma}{\mathrm{d}z_r},
\end{equation}
which describes the fraction of \pT{} inside the jet that is not contained within
the leading subjet~\cite{Neill:2021std}. 

\subsection{Subjet fragmentation in proton--proton collisions} \label{sec:results-pp}

\begin{figure}[!ht]
\centering{}
\includegraphics[scale=0.48]{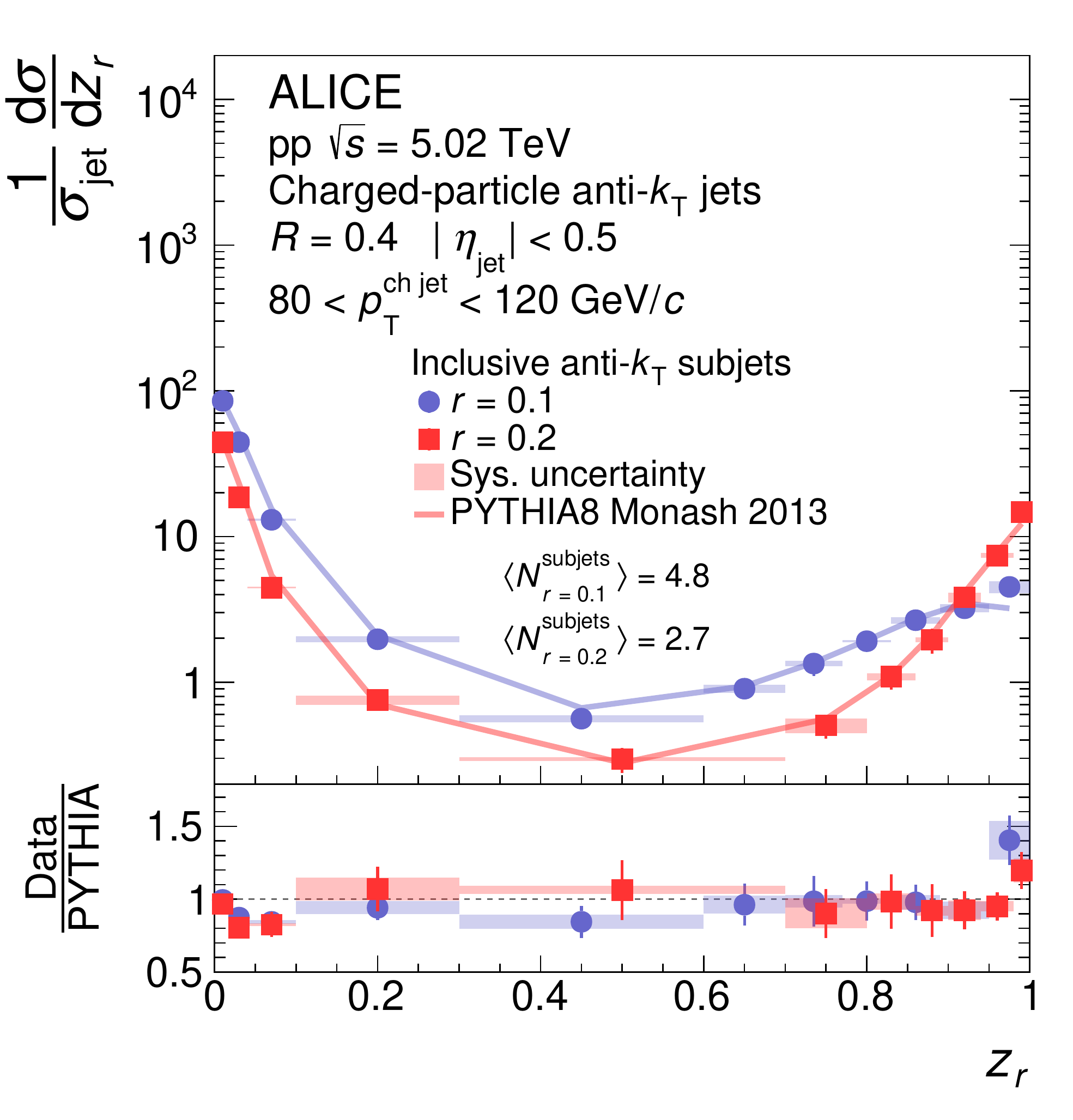}
\caption{ALICE measurements of inclusive subjet \zr{} distribution 
in \pp{} collisions for two different subjet radii, compared to PYTHIA8~\cite{Sjostrand:2014zea, Skands:2014pea}.}
\label{fig:zr-pp-inclusive}
\end{figure}

\begin{figure}[!hb]
\centering{}
\includegraphics[scale=0.48]{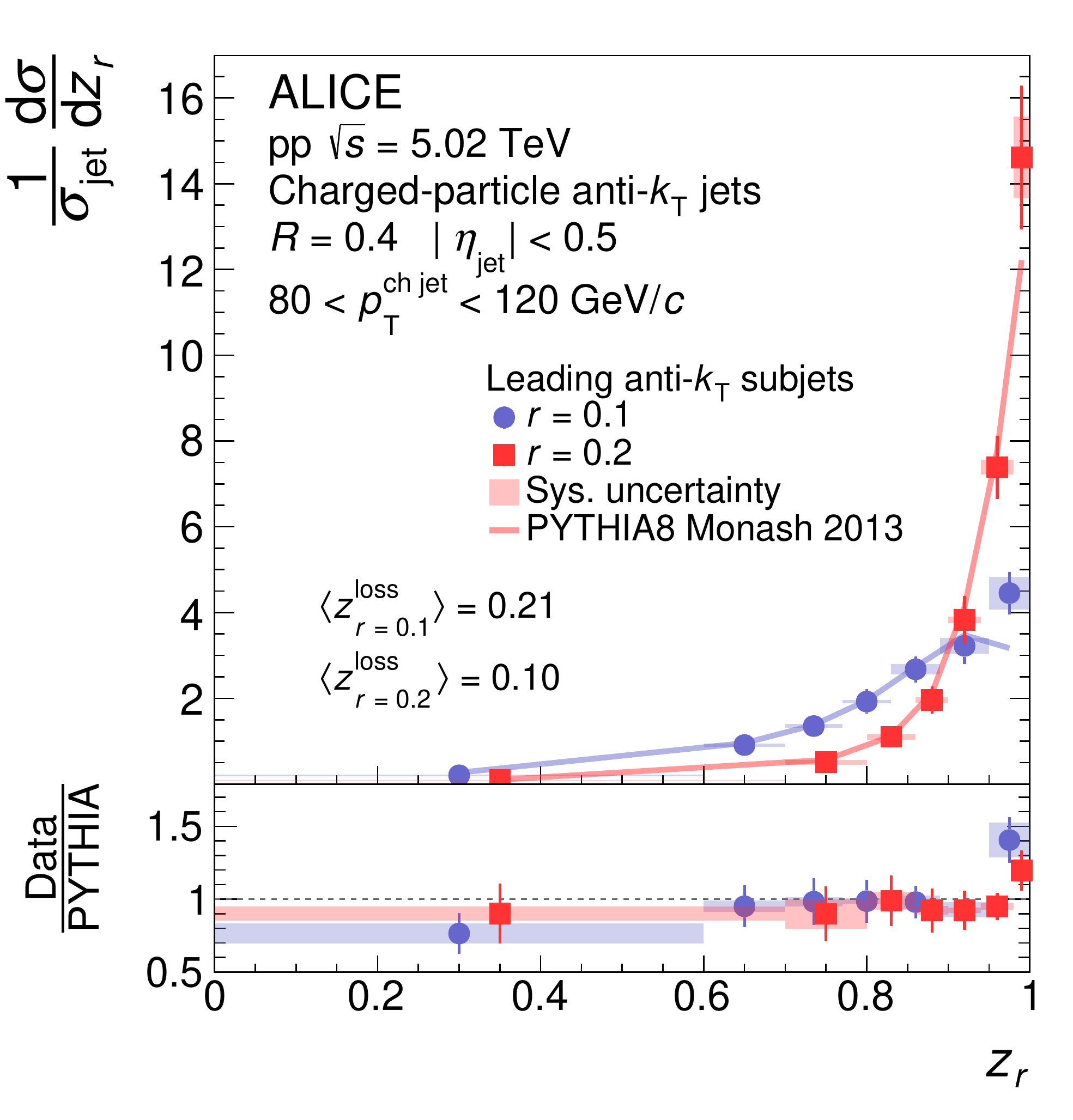}
\caption{ALICE measurements of leading subjet \zr{} distribution 
in \pp{} collisions for two different subjet radii, compared to PYTHIA8~\cite{Sjostrand:2014zea, Skands:2014pea}.}
\label{fig:zr-pp-leading}
\end{figure}

Figures~\ref{fig:zr-pp-inclusive} and~\ref{fig:zr-pp-leading} show the measured \zr{} distributions in \pp{} collisions at $\sqrt{s}=5.02\;\rm{TeV}$ for 
inclusive and leading subjets, respectively. 
For $\zr>0.5$ the leading and inclusive subjet distributions are identical, as expected.
In this region, the amplitude of the \zr{} distribution increases with \zr{}, 
with a more pronounced peak at large \zr{} for $r=0.2$ than for $r=0.1$ since larger subjets are more likely to capture
a larger fraction of the jet energy.
It is expected that as $\zr\rightarrow1$, the distributions will eventually decrease 
due to the increased splitting probability of soft emissions~\cite{Neill:2021std}.
This is, however, not visible in the data due to the coarseness of the bin sizes.
As \zr{} becomes small, the inclusive subjet distribution grows due to soft radiations emitted
from the leading subjet, whereas the leading subjet distribution falls to zero.
The fraction of \pT{} inside the jet that is not contained within
the leading subjet is $\langle z_{\rm{loss}}\rangle = 0.21$ for $r=0.1$
and decreases to $\langle z_{\rm{loss}}\rangle = 0.10$ for $r=0.2$.

The distributions are generally well described by PYTHIA8~\cite{Sjostrand:2014zea, Skands:2014pea}, however
some tension is observed in the largest \zr{} bin.
This may be due to threshold logarithms of $1-z$, which
may contribute significantly at all orders in the strong coupling constant $\alpha_s$ and are not directly included in PYTHIA8~\cite{Neill:2021std}.
In addition, hadronization effects are expected to be significant at large \zr{},
since hadronization causes a smearing of the fragmentation across
the boundary of the subjet and away from $\zr=1$.
However, due to ill-defined perturbative accuracy in general-purpose MC generators
such as PYTHIA
and the fact that they are tuned to reproduce data, it is difficult to draw detailed
physics conclusions from their comparison to data. 
In order to study these effects in greater detail, we
turn our attention to comparisons with analytical calculations based on perturbative QCD (pQCD).

Theoretical calculations of the \zr{} distribution in \pp{} collisions  have been carried out
within the Soft-Collinear Effective Theory (SCET) framework~\cite{SCETplaceholder}
for both inclusive and leading subjets for a variety of $r$ and $R$ values~\cite{Kang:2017mda, Neill:2021std}.
These calculations include all-order resummations of large logarithms of the jet radius and threshold logarithms to next-to-leading logarithmic (NLL$^\prime$) accuracy.
In order to compare these predictions to our measurement using
charged-particle jets, a ``forward folding'' procedure based on MC event generators is applied to account
for the fact that we measure only the charged component of jets~\cite{ALICE:2021njq}.
Although the calculations are provided with hadronization corrections included~\cite{Neill:2021std}, we additionally applied 
a bin-by-bin correction to account for multi-parton interactions using the procedure outlined in Ref.~\cite{ALICE:2021njq}.

\begin{figure}[!th]
\centering
\hspace*{-1cm}\includegraphics[scale=0.75]{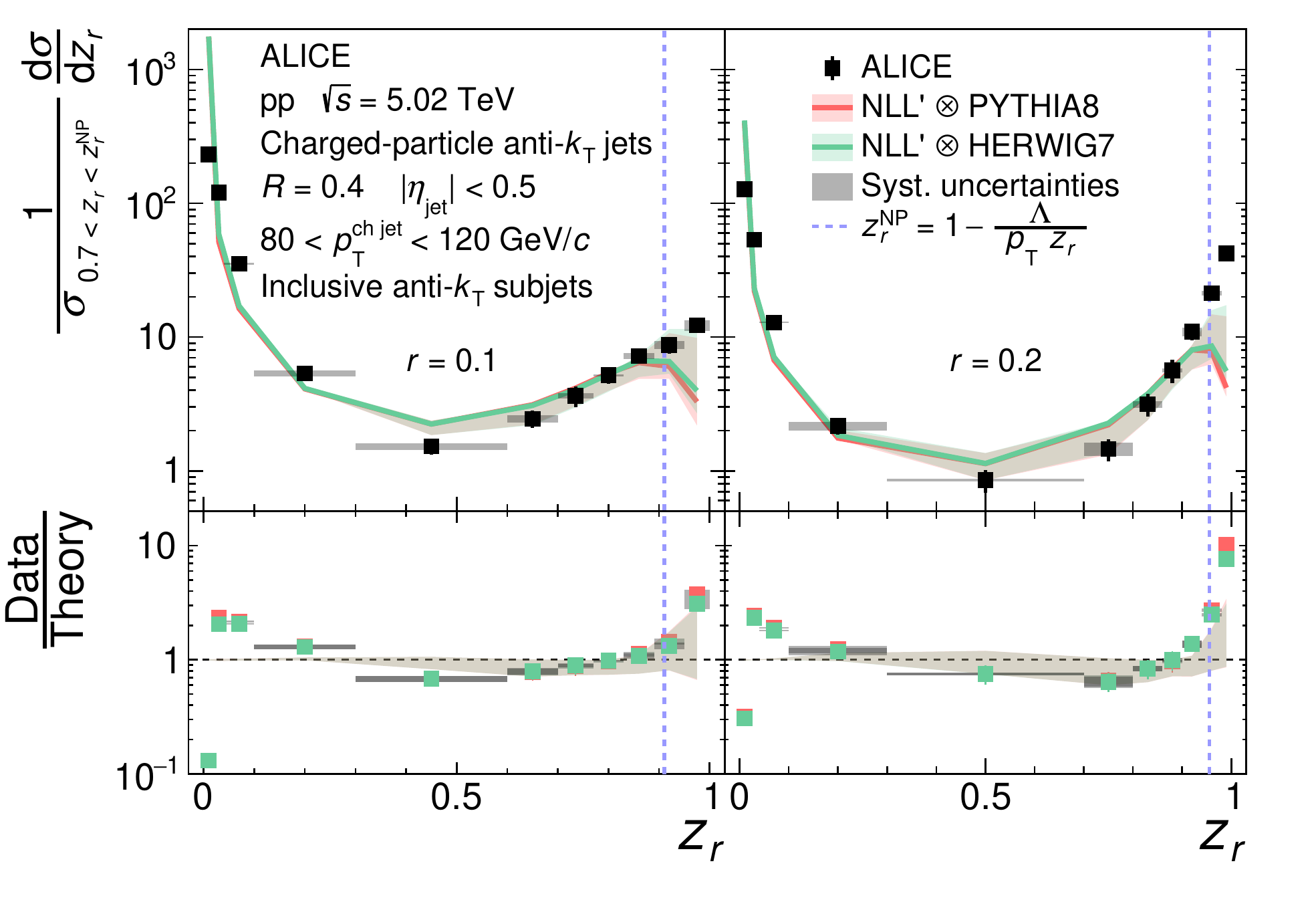}\vspace{-0.5cm}
\hspace*{-1cm}\includegraphics[scale=0.75]{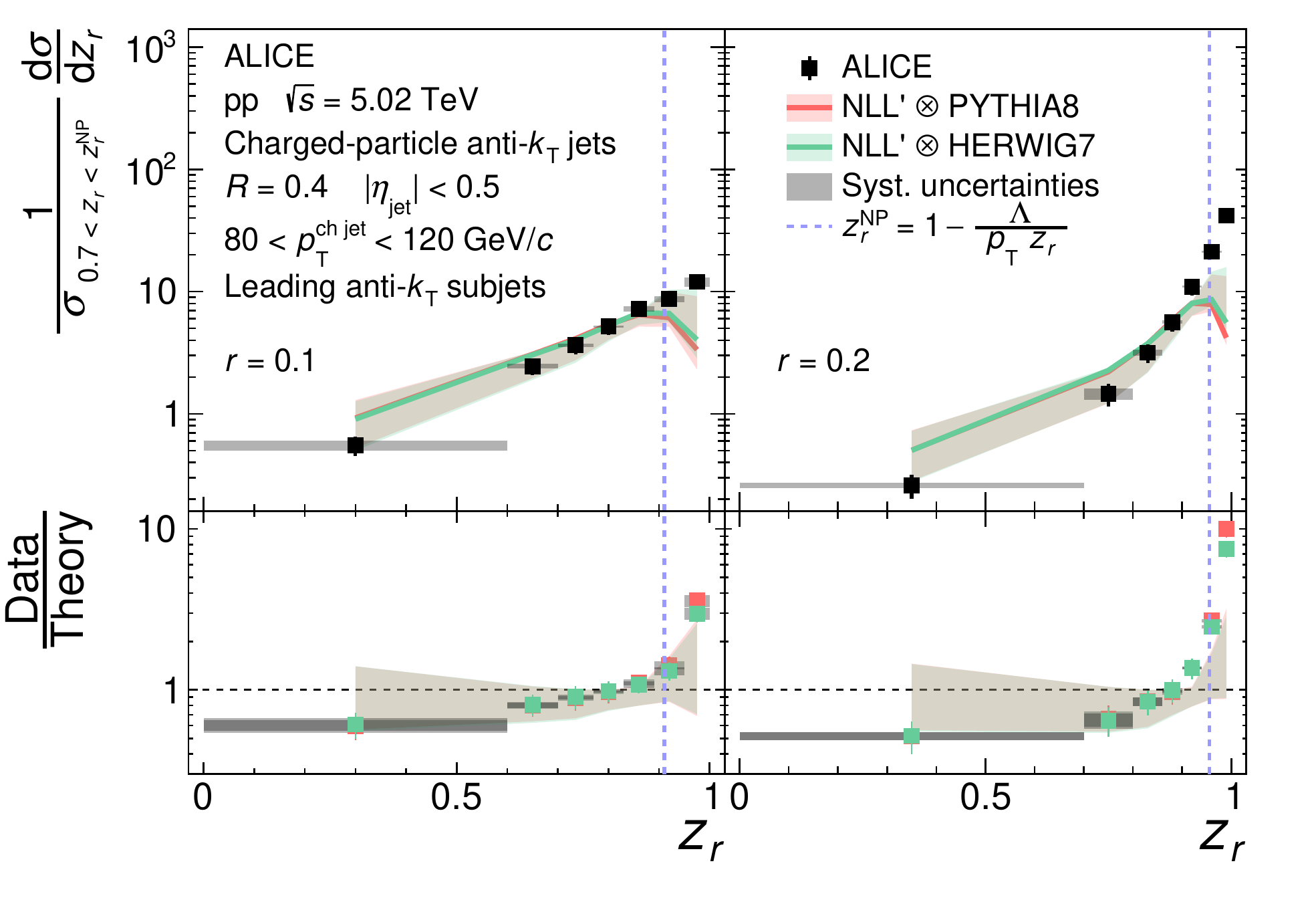}\vspace{-0.5cm}
\caption{ALICE measurements of inclusive (top) and leading (bottom) subjet \zr{} distributions in \pp{} collisions at $\sqrt{s}=5.02$ TeV, compared to NLL$^\prime$ predictions carried out with SCET~\cite{Kang:2017mda,Neill:2021std} and
corrected for missing neutral-particle energy and multi-parton interaction effects using PYTHIA8~\cite{Sjostrand:2014zea}
or HERWIG7~\cite{Bellm:2015jjp}. The shaded bands denote systematic uncertainty on the NLL$^\prime$ calculations. The distributions are normalized such that the
integral of the region defined by $0.7<\zr<\zrNP$ is unity,
where \zrNP{} is denoted by the dashed vertical blue lines. 
The non-perturbative scale in Eq.~\ref{eq:8} is taken to be $\Lambda=1\;\GeVc$.
In determining the normalization, bins that overlap with the dashed blue line are considered to be in the non-perturbative (right) region.}
\label{fig:zr-theory}
\end{figure}

Figure~\ref{fig:zr-theory} compares the measured \zr{} distributions to NLL$^\prime$
calculations for inclusive and leading subjets~\cite{Kang:2017mda, Neill:2021std}. 
Two sets of NLL$^\prime$ results are reported in the figure, which are obtained using
either PYTHIA8~\cite{Sjostrand:2014zea} or HERWIG7~\cite{Bellm:2015jjp} to account
for charged-particle corrections, which show generally similar behavior.
Uncertainties on the analytical predictions were estimated in Refs.~\cite{Kang:2017mda, Neill:2021std} by
varying the combinations of scales that emerge in the calculation.
The softest of these scales determines a transition between the 
perturbative and non-perturbative regimes:
\begin{equation} \label{eq:8}
\zrNP \approx 1 - \left( \frac{\Lambda}{\pT r} \right),
\end{equation}
where $\Lambda$ is the energy scale at which $\alpha_s$ becomes non-perturbative.
To denote this transition, we draw a dashed vertical blue line 
at $\Lambda=1\;\GeVc$, taking \pT{} to be the weighted average \pTchjet{} in
the considered interval scaled by 120\% to approximately translate the \pT{}
scale from charged-particle jets to full jets.
While we draw the line at a discrete value in order to provide guidance, we remind the reader that 
the transition from values of \zr{} that are dominated by perturbative versus
non-perturbative physics is actually smooth. 
Note that we display the non-perturbative transition only at large \zr{}, 
although a similar transition occurs at small \zr{} which is not addressed in this study~\cite{Neill:2020bwv, Neill:2020tzl}.
In the results shown in Fig~\ref{fig:zr-theory}, the cross section is scaled according to the integral of the distribution
in a subset of the perturbative region,

\begin{equation} \label{eq:9}
\frac{1}{\sigma_{0.7<\zr<\zrNP}} \frac{\mathrm{d}\sigma}{\mathrm{d}\zr},
\qquad\mathrm{where}\qquad 
\sigma_{0.7<\zr<\zrNP} = \int_{0.7}^{\zrNP} \frac{\mathrm{d}\sigma}{\mathrm{d}\zr}\mathrm{d}\zr.
\end{equation}

In the inclusive subjet case, the measured \zr{} distributions are generally in agreement with the SCET
calculations within uncertainties in the intermediate region $0.1\lesssim \zr \lesssim 0.9$.
The calculations begin to diverge from the data at large \zr{} in the non-perturbative regime, where the theoretical calculations are expected to break down. For $r=0.1$, the calculations can, in fact, describe the data at large \zr{} within the systematic uncertainties of the calculation, whereas for $r=0.2$ this is no longer the case. 
At small \zr{}, the calculations diverge from the data, with a large overestimate of the peak at $\zr<0.02$ relative to the magnitude of the theoretical uncertainties.
The calculations do not include a resummation of large logarithms of small \zr{}, which suggests that such a resummation is needed to describe the data. The measured distributions can serve to test future calculations that include small \zr{} resummation,
which is relevant for attempts to calculate hadron observables
using perturbatively calculable jet functions in the $r\rightarrow0$ limit~\cite{Neill:2020bwv, Neill:2020tzl}.

In the leading subjet case, we observe identical behavior to the inclusive subjet case in the region $\zr>0.5$, since the distributions coincide.
For $\zr<0.5$, where the inclusive and leading distributions differ, 
the calculations involve nonlinear evolution of the jet fragmentation function.
In this region, the NLL$^\prime$ results are consistent with the data within the uncertainties of the calculation. Note that this comparison involves only a single bin due to the fact that the \zr{} distributions fall steeply as \zr{} decreases. More differential data would allow for stricter tests of the non-linear evolution of leading jet fragmentation functions. 
A test of related calculations of leading dijet energy spectra have recently been examined in e$^+$e$^-$ collisions~\cite{Chen:2021uws}, finding good agreement.
Finally, we note that the quantity $\langle z_{\rm{loss}}\rangle$  
described in Eq.~\ref{eq:3} is strongly affected by the non-perturbative region $\zr>\zrNP$,
since a significant fraction of the integral is located in this interval. This presents a challenge to the prospects for theoretically calculating $\langle z_{\rm{loss}}\rangle$, and calls for further theoretical studies.

\subsection{Subjet fragmentation in Pb--Pb collisions}

\begin{figure}[!ht]
\centering{}
\includegraphics[scale=0.47]{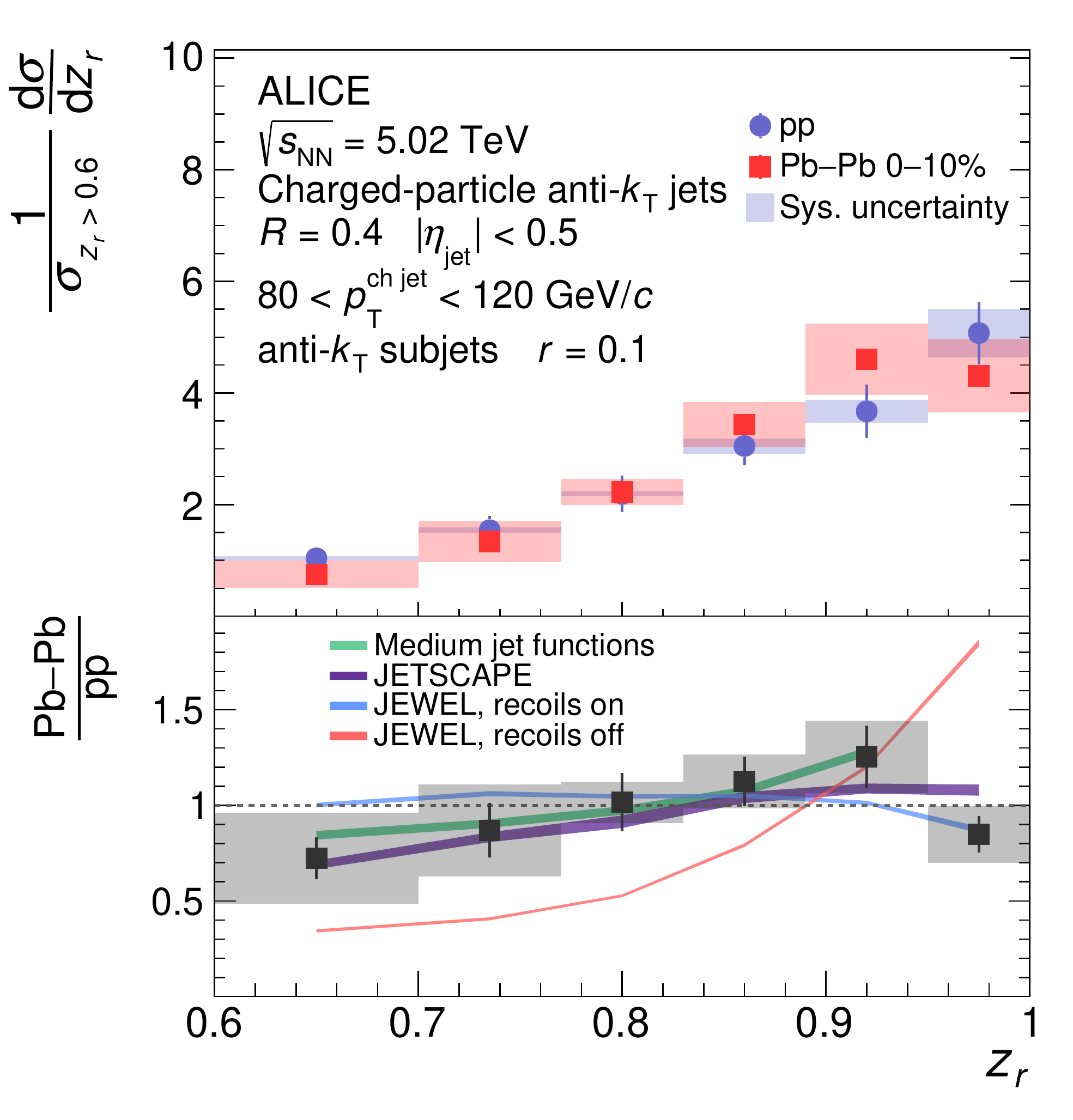}
\caption{Measurements of subjet \zr{} distributions for subjet radius $r=0.1$ in \pp{} and 0--10\% central \PbPb{} collisions.
The bottom panel displays the ratio of the distributions in \PbPb{} to \pp{} collisions, 
along with comparison to theoretical predictions~\cite{Kang:2017mda, Qiu:2019sfj, Putschke:2019yrg, He:2015pra, Majumder:2013re, Zapp:2012ak, Zapp:2013vla}.}
\label{fig:zr-AA-r01}
\end{figure}

\begin{figure}[!hb]
\centering{}
\includegraphics[scale=0.47]{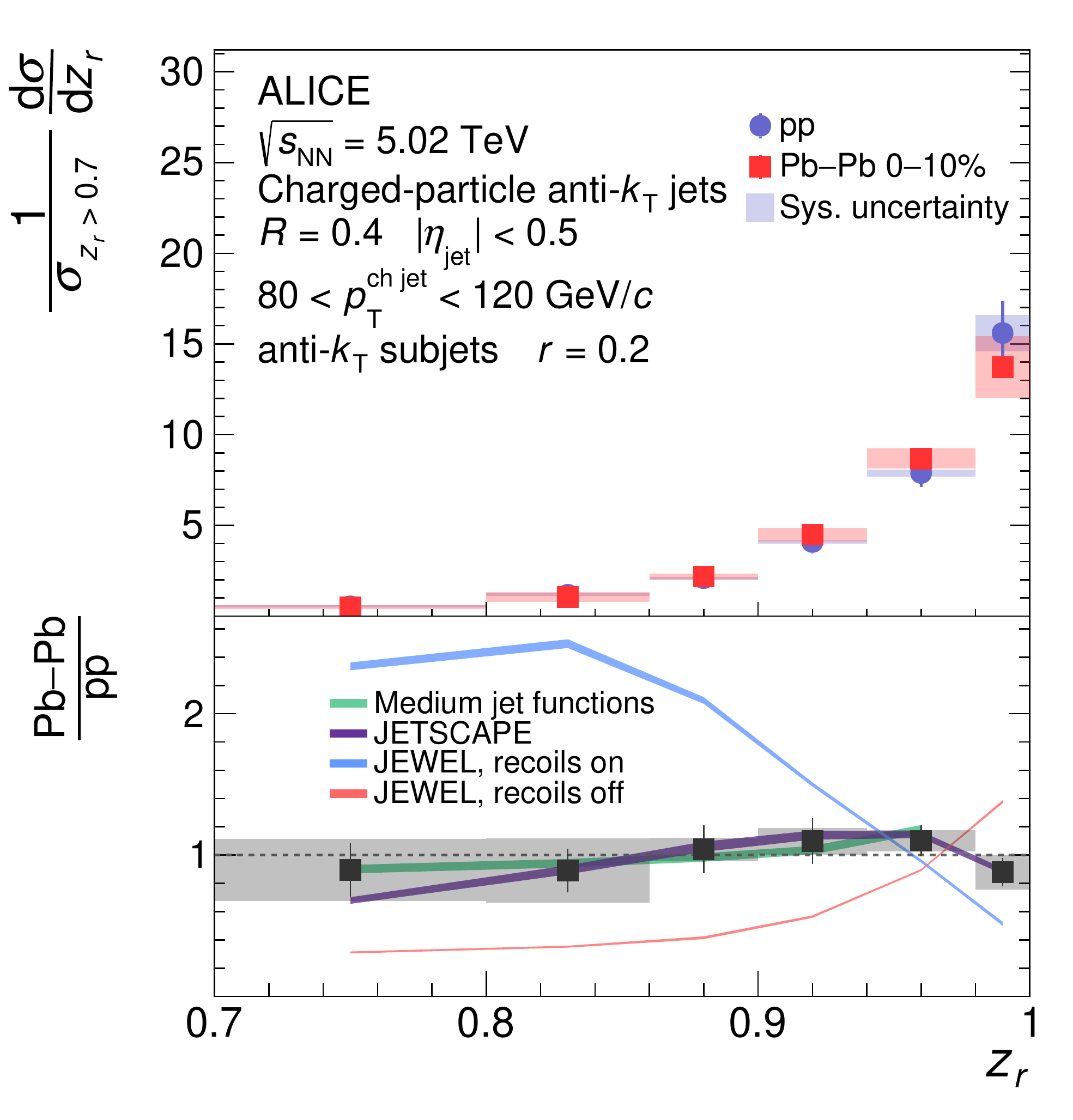}
\caption{Measurements of subjet \zr{} distributions for subjet radius $r=0.2$ in \pp{} and 0--10\% central \PbPb{} collisions. 
The bottom panel displays the ratio of the distributions in \PbPb{} to \pp{} collisions, 
along with comparison to theoretical predictions~\cite{Kang:2017mda, Qiu:2019sfj, Putschke:2019yrg, He:2015pra, Majumder:2013re, Zapp:2012ak, Zapp:2013vla}.}
\label{fig:zr-AA-r02}
\end{figure}

Figures~\ref{fig:zr-AA-r01} and~\ref{fig:zr-AA-r02} show the leading subjet \zr{} distributions 
in \pp{} and 0--10\% central \PbPb{} collisions for $r=0.1$ and $r=0.2$, respectively,
with their ratios displayed in the bottom panels.
We report a restricted range in \zr{} in \PbPb{} collisions due to the contamination from the underlying 
event at low-\zr{}, as explained in Sec.~\ref{sec:analysis}, although we note that this excludes only a small portion of the leading subjet distribution.
The reported distributions are accordingly normalized to the cross section
of inclusive charged-particle jets conditioned with \zr{} in the reported range.
The relative uncertainties are assumed to be uncorrelated between \pp{} and \PbPb{} collisions, 
and are added in quadrature in the ratio.

For both $r=0.1$ and $r=0.2$, the distributions are consistent with no modification of 
the \zr{} distribution in central \PbPb{} compared to \pp{} collisions.
However, the distributions are also consistent with a hardening effect
in \PbPb{} compared to \pp{} collisions that reverses as $\zr \rightarrow 1$.
For $r=0.1$, we observe consistency with stronger hardening effects
than for $r=0.2$.
To understand the behavior of the \zr{} distribution, we note that in vacuum there are significant
differences in the parton-to-subjet fragmentation functions 
between quarks and gluons, with the fraction of quark-initiated jets
increasing with \zr{}~\cite{Neill:2021std}.
If the QGP suppresses gluon jets more than quark jets,
then a hardening of the \zr{} distribution is expected
– in line with previous measurements of hadron fragmentation~\cite{Spousta:2015fca}.
On the other hand, medium-induced radiations
will in general shift the distribution to smaller \zr{}.
The competition between these two distinct sources of jet substructure modification -- 
quark vs. gluon suppression and medium-induced radiation -- 
can result in non-trivial modification to the shape of the \zr{} distribution
in different intervals of \zr{}.
As $\zr \rightarrow 1$, the jet sample in vacuum
becomes almost entirely dominated by quark jets – thereby rendering the quark vs. gluon fraction
modification negligible.
This presents an opportunity to 
expose a region of quark-initiated jets depleted by soft medium-induced emissions.
Our measurements are qualitatively consistent with such a modification pattern: a hardening of the 
\zr{} distribution due to the relative suppression of gluon vs. quark initiated jets, 
followed by a turnover of the distribution as $\zr \rightarrow 1$ due to medium-induced soft radiations.

We compare the ratio of the measurements in pp and \PbPb{} collisions with several theoretical models of jet quenching:

\begin{itemize}[topsep=-4pt]

\item {\it Medium jet functions}~\cite{Kang:2017mda, Qiu:2019sfj}
are a SCET-based calculation obtained by modifying the \zr{} distributions from \pp{} collisions 
according to the medium-modified parton-to-jet fragmentation functions extracted in Ref.~\cite{Qiu:2019sfj}.
The quark/gluon fractions in the extracted medium-modified jet function 
exhibit a relative suppression factor of approximately four between gluon jets and quark jets.

\item {\it JETSCAPE}~\cite{Putschke:2019yrg, JETSCAPE:2022jer} 
consists of a medium-modified parton
shower calculated with the MATTER model~\cite{Majumder:2013re} controlling the high-virtuality 
phase and the Linear Boltzmann Transport (LBT) model describing the low-virtuality phase~\cite{He:2015pra}. The version of JETSCAPE used for this calculation employs a jet transport coefficient, $\hat{q}$, that includes dependence on parton virtuality, in addition to dependence on the local temperature and running of the parton-medium coupling.
The calculation includes medium recoil particles, and a subtraction of the thermal component of the recoils is performed by summing the transverse momentum of the thermal particles within the jet (subjet) radius and subtracting this from the corresponding jet (subjet) transverse momentum.

\item {\it JEWEL}~\cite{Zapp:2012ak, Zapp:2013vla}
implements BDMPS-based medium-induced gluon radiation in a medium
modeled with a Bjorken expansion. 
We use JEWEL 2.2.0 with an initial temperature $T_i=590$ MeV and initial quenching time $\tau_i=0.4$, 
which provides an accurate description of a variety of jet quenching observables~\cite{KunnawalkamElayavalli:2017hxo}.
The impact of medium recoil is studied by displaying results both with and without recoils enabled. 
In the case with recoils included, the thermal component of the recoils is subtracted with the same method used in the JETSCAPE calculation (which is similar to the ``4MomSub'' method~\cite{KunnawalkamElayavalli:2017hxo}) except randomly discarding 33\% of the thermal particles (which JEWEL assigns to be neutral) in order to account for the fact that our measurement uses charged-particle jets.

\end{itemize}

For the JETSCAPE and JEWEL simulations, the width of the curves denotes statistical uncertainty. For the ``Medium jet functions'' calculation,
systematic uncertainties are included but are smaller than the width of the plotted curve.

The comparison of our result to the ``Medium jet functions'' calculation
provides a test of universality of jet fragmentation functions in the QGP, 
since the calculation uses a parton-to-jet function extracted from inclusive jet measurements,
and employs it as the parton-to-subjet function in the \zr{} calculation.
We find a consistent description of the \zr{} distribution, 
and therefore consistency with the universality of jet fragmentation in the QGP.
While this does not exclude process-dependent effects or factorization breaking,
it does place constraints on the magnitude of such effects, and 
establishes a new avenue to search for them.
These measurements can be used to directly extract the parton-to-subjet function in future work
and serve as input for global tests of factorization breaking in the QGP (see Ref.~\cite{Qiu:2019sfj}).

The JETSCAPE model describes the data well within the precision of our measurement. The JEWEL model, on the other hand, describes the data 
well for $r=0.1$ when recoils are included, 
but fails to describe the data for $r=0.2$ or when recoils are not included. 
For both $r=0.1$ and $r=0.2$, there are large differences in the JEWEL predictions depending whether recoils are enabled, suggesting that 
this observable may be significantly impacted by medium response. 
In general, it is expected that medium response will soften the \zr{} distribution since it tends to broaden reconstructed jets.
We indeed observe this in the results of the JEWEL calculations, where the \zr{} distribution in the largest \zr{} bin is significantly suppressed when recoils are included compared to when recoils are disabled.
However, it appears that for $r=0.2$, this suppression is significantly stronger than the experimental data allows (noting that the large enhancement observed at smaller \zr{} is necessitated by the suppression at large \zr{} due to the self-normalization condition). 
This corroborates previous observations that the medium response implementation in JEWEL, which does not include rescattering of the medium response particles, overestimates the impact of medium recoil, but that the calculations with and without recoil generally bracket the experimental data (see e.g. Ref.~\cite{CMS:2021vui,ALICE:2017nij}).

\begin{figure}[!t]
\centering{}
\includegraphics[scale=0.55]{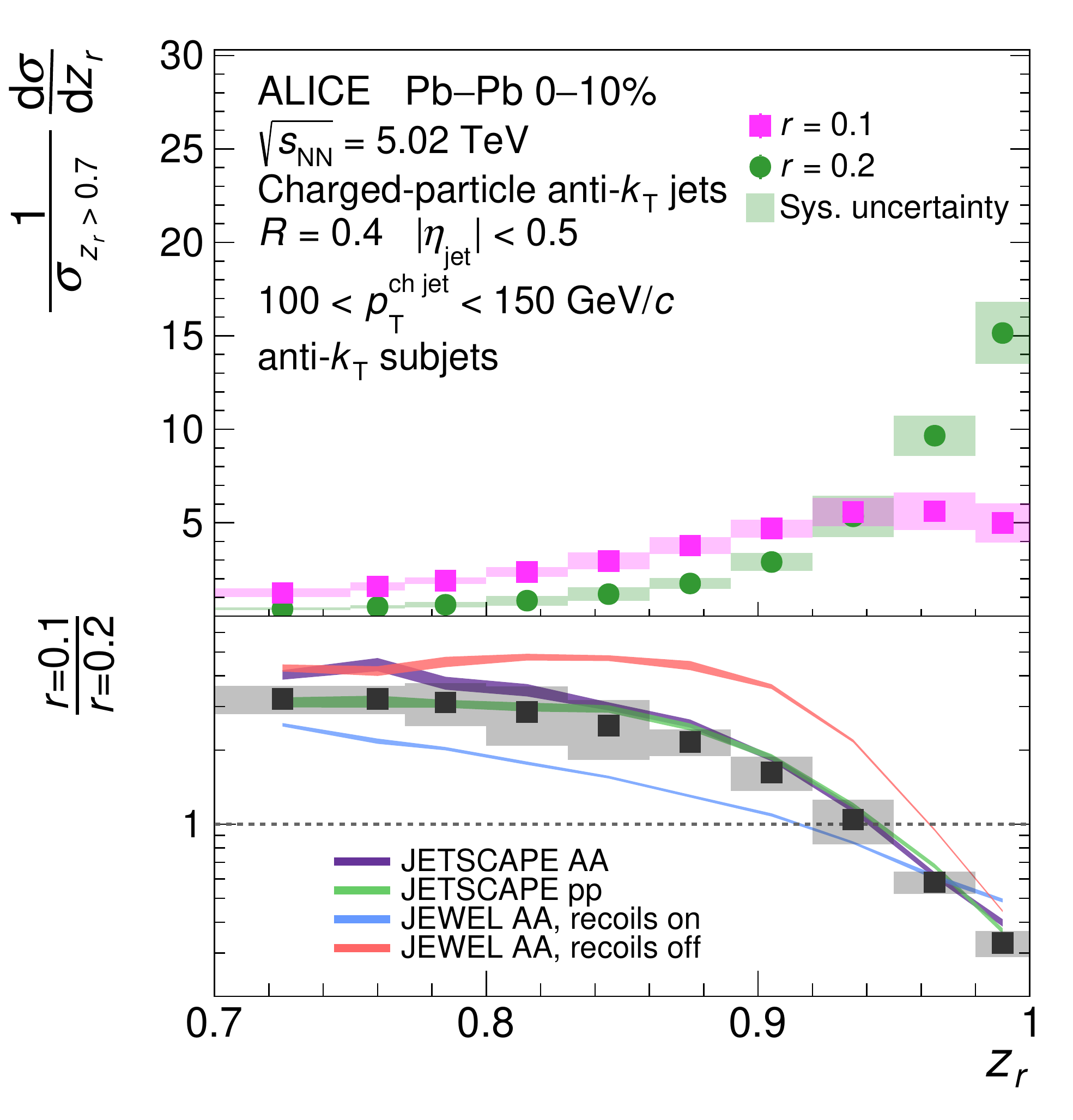}
\caption{Measurements of subjet \zr{} distributions for subjet radii $r=0.1$ 
and $r=0.2$ in 0--10\% central \PbPb{} collisions.
The bottom panel displays the ratio of the distributions for $r=0.1$ to $r=0.2$, 
along with comparison to JEWEL and JETSCAPE model predictions~\cite{Putschke:2019yrg, He:2015pra, Majumder:2013re, Zapp:2012ak, Zapp:2013vla}.}
\label{fig:zr-AA-ratio}
\end{figure}

In order to test the $r$-dependence of the \zr{} distribution with higher precision, 
in Fig.~\ref{fig:zr-AA-ratio} we compare the \zr{} distributions measured in 0--10\% central \PbPb{} 
collisions with $r=0.1$ and $r=0.2$ for the \pTchjet{} interval between 100 and 150 GeV/$c$.
The ratio of the two distributions is shown in the bottom panel of Fig.~\ref{fig:zr-AA-ratio}.
Since the two measurements are correlated, 
the systematic uncertainty of the ratio partially cancels out.
While separate values of $\sigma_{\zr>0.7}$ are used to normalize
the $r=0.1$ and $r=0.2$ distributions, the normalization factors only differ by the
integral of the $\zr<0.7$ tails and are therefore within a few percent.
We refrain from constructing the corresponding \pp{} ratio due to sizeable  
statistical uncertainties of the recorded \pp{} data set.

We compare the $\frac{r=0.1}{r=0.2}$ ratio to the JETSCAPE and JEWEL models discussed above.
We find that JEWEL fails to describe the ratio either with recoils on or recoils off, but that the two implementations bracket most of the data.
However, the JEWEL calculations predict that at large \zr{}, the $r=0.1$ ``core'' of the jet contains 
more \pT{} relative to $r=0.2$ as compared to the experimental data -- i.e. that the large-\zr{} 
fragmentation is narrower in JEWEL than in experimental data.

The JETSCAPE model describes the ratio significantly better, 
however given the precision of the measurement, 
we find significant tension in the shape of the distribution with the JETSCAPE prediction.
While this may not be immediately obvious by eye, 
we note that due to the self-normalization condition in the \zr{} distributions
in the top panel of Fig.~\ref{fig:zr-AA-ratio}, a shift in one point necessitates
an opposite shift in the remaining points to ensure the distribution integrates to unity.
For example, if the value of the ratio in the 
rightmost bin $0.98<\zr<1$ of the JETSCAPE calculation were to move down,
the leftmost bins near $0.7<\zr<0.8$ would have to compensate by moving up, 
rendering the calculation unable to reproduce the experimental data.
The distribution from JETSCAPE simulations for \pp{} collisions, on the other hand, describes the $\frac{r=0.1}{r=0.2}$ ratio well; however, one must use caution in interpreting this,
since in Sec.~\ref{sec:results-pp} we have discussed challenges in achieving an accurate description of the \pp{} baseline at large \zr{}.
The JETSCAPE calculation exhibits a hint that the $r=0.1$ ``core'' of the jet contains 
more \pT{} relative to $r=0.2$ as compared to the experimental data,
similar to JEWEL although with less significant tension.

\section{Conclusions}

We have presented new measurements of subjet fragmentation with ALICE. 
In \pp{} collisions,
we find agreement of pQCD calculations with the data in the perturbative regime at intermediate \zr{} and discrepancies at large \zr{} which may imply 
that threshold resummation and hadronization play important roles as the distribution becomes increasingly non-perturbative. 
The PYTHIA8 event generator generally describes the data well, however some tension is observed at large \zr{}, which is consistent with these findings given that threshold resummation is not directly included in PYTHIA8.
In the inclusive subjet case, we find a disagreement of the pQCD calculations with the data at small \zr{}, suggesting a need to include a small-\zr{} resummation in order to describe the data.
These measurements provide future opportunities to study threshold and small-\zr{} resummations, and motivate new measurements extended to even smaller values of $r$, which are relevant for understanding parton-hadron duality and the interplay of jet observables and hadron observables as $r\rightarrow0$~\cite{Neill:2020bwv, Neill:2020tzl}.

In heavy-ion collisions, these measurements serve as a key ingredient to study the high-$z$ region of 
jet quenching and test the universality of jet fragmentation in the QGP. 
By comparing the \zr{} distributions for $r=0.1$ and $r=0.2$ to Monte Carlo jet quenching models, we find indications that quenched jets at large \zr{} are
narrower in JEWEL and JETSCAPE than in experimental data.
By probing large \zr{}, these measurements can isolate a region of quark-dominated 
jets with an inclusive jet sample alone,
offering the potential to expose a sample of jets depleted by medium-induced soft radiation.
Together, our measurements demonstrate that while the large-\zr{} region is theoretically challenging 
to describe in \pp{} collisions
due to threshold resummation and hadronization effects, it is a particularly interesting
region to study jet modification in heavy-ion collisions.
This calls for theoretical investigation of the large-\zr{} region in greater detail.
Future measurements of \zr{} in coincidence with other substructure observables
such as the groomed jet radius~\cite{ALICE:2021obz} offer the potential to disentangle
medium-induced soft radiation effects from differences in the suppression of gluon vs. quark jets.
By comparing our measurements to perturbative calculations based on QCD factorization, 
we find consistency with the universality of jet fragmentation
and no indication of factorization breaking in the QGP.
These measurements can be used as input to extract the parton-to-subjet fragmentation function in future work
and perform global tests of factorization breaking in the QGP.

\newenvironment{acknowledgement}{\relax}{\relax}
\begin{acknowledgement}
\section*{Acknowledgements}
We gratefully acknowledge Duff Neill, Felix Ringer, Nobuo Sato, and the JETSCAPE Collaboration
for providing theoretical predictions.

The ALICE Collaboration would like to thank all its engineers and technicians for their invaluable contributions to the construction of the experiment and the CERN accelerator teams for the outstanding performance of the LHC complex.
The ALICE Collaboration gratefully acknowledges the resources and support provided by all Grid centres and the Worldwide LHC Computing Grid (WLCG) collaboration.
The ALICE Collaboration acknowledges the following funding agencies for their support in building and running the ALICE detector:
A. I. Alikhanyan National Science Laboratory (Yerevan Physics Institute) Foundation (ANSL), State Committee of Science and World Federation of Scientists (WFS), Armenia;
Austrian Academy of Sciences, Austrian Science Fund (FWF): [M 2467-N36] and Nationalstiftung f\"{u}r Forschung, Technologie und Entwicklung, Austria;
Ministry of Communications and High Technologies, National Nuclear Research Center, Azerbaijan;
Conselho Nacional de Desenvolvimento Cient\'{\i}fico e Tecnol\'{o}gico (CNPq), Financiadora de Estudos e Projetos (Finep), Funda\c{c}\~{a}o de Amparo \`{a} Pesquisa do Estado de S\~{a}o Paulo (FAPESP) and Universidade Federal do Rio Grande do Sul (UFRGS), Brazil;
Bulgarian Ministry of Education and Science, within the National Roadmap for Research Infrastructures 2020¿2027 (object CERN), Bulgaria;
Ministry of Education of China (MOEC) , Ministry of Science \& Technology of China (MSTC) and National Natural Science Foundation of China (NSFC), China;
Ministry of Science and Education and Croatian Science Foundation, Croatia;
Centro de Aplicaciones Tecnol\'{o}gicas y Desarrollo Nuclear (CEADEN), Cubaenerg\'{\i}a, Cuba;
Ministry of Education, Youth and Sports of the Czech Republic, Czech Republic;
The Danish Council for Independent Research | Natural Sciences, the VILLUM FONDEN and Danish National Research Foundation (DNRF), Denmark;
Helsinki Institute of Physics (HIP), Finland;
Commissariat \`{a} l'Energie Atomique (CEA) and Institut National de Physique Nucl\'{e}aire et de Physique des Particules (IN2P3) and Centre National de la Recherche Scientifique (CNRS), France;
Bundesministerium f\"{u}r Bildung und Forschung (BMBF) and GSI Helmholtzzentrum f\"{u}r Schwerionenforschung GmbH, Germany;
General Secretariat for Research and Technology, Ministry of Education, Research and Religions, Greece;
National Research, Development and Innovation Office, Hungary;
Department of Atomic Energy Government of India (DAE), Department of Science and Technology, Government of India (DST), University Grants Commission, Government of India (UGC) and Council of Scientific and Industrial Research (CSIR), India;
National Research and Innovation Agency - BRIN, Indonesia;
Istituto Nazionale di Fisica Nucleare (INFN), Italy;
Japanese Ministry of Education, Culture, Sports, Science and Technology (MEXT) and Japan Society for the Promotion of Science (JSPS) KAKENHI, Japan;
Consejo Nacional de Ciencia (CONACYT) y Tecnolog\'{i}a, through Fondo de Cooperaci\'{o}n Internacional en Ciencia y Tecnolog\'{i}a (FONCICYT) and Direcci\'{o}n General de Asuntos del Personal Academico (DGAPA), Mexico;
Nederlandse Organisatie voor Wetenschappelijk Onderzoek (NWO), Netherlands;
The Research Council of Norway, Norway;
Commission on Science and Technology for Sustainable Development in the South (COMSATS), Pakistan;
Pontificia Universidad Cat\'{o}lica del Per\'{u}, Peru;
Ministry of Education and Science, National Science Centre and WUT ID-UB, Poland;
Korea Institute of Science and Technology Information and National Research Foundation of Korea (NRF), Republic of Korea;
Ministry of Education and Scientific Research, Institute of Atomic Physics, Ministry of Research and Innovation and Institute of Atomic Physics and University Politehnica of Bucharest, Romania;
Ministry of Education, Science, Research and Sport of the Slovak Republic, Slovakia;
National Research Foundation of South Africa, South Africa;
Swedish Research Council (VR) and Knut \& Alice Wallenberg Foundation (KAW), Sweden;
European Organization for Nuclear Research, Switzerland;
Suranaree University of Technology (SUT), National Science and Technology Development Agency (NSTDA), Thailand Science Research and Innovation (TSRI) and National Science, Research and Innovation Fund (NSRF), Thailand;
Turkish Energy, Nuclear and Mineral Research Agency (TENMAK), Turkey;
National Academy of  Sciences of Ukraine, Ukraine;
Science and Technology Facilities Council (STFC), United Kingdom;
National Science Foundation of the United States of America (NSF) and United States Department of Energy, Office of Nuclear Physics (DOE NP), United States of America.
In addition, individual groups or members have received support from:
Marie Sk\l{}odowska Curie, Strong 2020 - Horizon 2020, European Research Council (grant nos. 824093, 896850, 950692), European Union;
Academy of Finland (Center of Excellence in Quark Matter) (grant nos. 346327, 346328), Finland;
Programa de Apoyos para la Superaci\'{o}n del Personal Acad\'{e}mico, UNAM, Mexico.

\end{acknowledgement}

\bibliographystyle{utphys}
\bibliography{main.bib}

\newpage
\appendix

%
%

\section{The ALICE Collaboration}
\label{app:collab}
\begin{flushleft} 
\small

S.~Acharya\,\orcidlink{0000-0002-9213-5329}\,$^{\rm 124,131}$, 
D.~Adamov\'{a}\,\orcidlink{0000-0002-0504-7428}\,$^{\rm 86}$, 
A.~Adler$^{\rm 69}$, 
G.~Aglieri Rinella\,\orcidlink{0000-0002-9611-3696}\,$^{\rm 32}$, 
M.~Agnello\,\orcidlink{0000-0002-0760-5075}\,$^{\rm 29}$, 
N.~Agrawal\,\orcidlink{0000-0003-0348-9836}\,$^{\rm 50}$, 
Z.~Ahammed\,\orcidlink{0000-0001-5241-7412}\,$^{\rm 131}$, 
S.~Ahmad\,\orcidlink{0000-0003-0497-5705}\,$^{\rm 15}$, 
S.U.~Ahn\,\orcidlink{0000-0001-8847-489X}\,$^{\rm 70}$, 
I.~Ahuja\,\orcidlink{0000-0002-4417-1392}\,$^{\rm 37}$
% ...

\section*{Affiliation Notes}

$^{\rm I}$ Deceased\\
$^{\rm II}$ Also at: Max-Planck-Institut f\"{u}r Physik, Munich, Germany\\
$^{\rm III}$ Also at: Italian National Agency for New Technologies, Energy and Sustainable Economic Development (ENEA), Bologna, Italy\\
$^{\rm IV}$ Also at: Dipartimento DET del Politecnico di Torino, Turin, Italy\\
$^{\rm V}$ Also at: Department of Applied Physics, Aligarh Muslim University, Aligarh, India\\
$^{\rm VI}$ Also at: Institute of Theoretical Physics, University of Wroclaw, Poland\\
$^{\rm VII}$ Also at: An institution covered by a cooperation agreement with CERN\\

\section*{Collaboration Institutes}

$^{1}$ A.I. Alikhanyan National Science Laboratory (Yerevan Physics Institute) Foundation, Yerevan, Armenia\\
$^{2}$ AGH University of Science and Technology, Cracow, Poland\\
$^{3}$ Bogolyubov Institute for Theoretical Physics, National Academy of Sciences of Ukraine, Kiev, Ukraine\\
$^{4}$ Bose Institute, Department of Physics  and Centre for Astroparticle Physics and Space Science (CAPSS), Kolkata, India\\
$^{5}$ California Polytechnic State University, San Luis Obispo, California, United States\\
$^{6}$ Central China Normal University, Wuhan, China\\
$^{7}$ Centro de Aplicaciones Tecnol\'{o}gicas y Desarrollo Nuclear (CEADEN), Havana, Cuba\\
$^{8}$ Centro de Investigaci\'{o}n y de Estudios Avanzados (CINVESTAV), Mexico City and M\'{e}rida, Mexico\\
$^{9}$ Chicago State University, Chicago, Illinois, United States\\
$^{10}$ China Institute of Atomic Energy, Beijing, China\\
$^{11}$ Chungbuk National University, Cheongju, Republic of Korea\\
$^{12}$ Comenius University Bratislava, Faculty of Mathematics, Physics and Informatics, Bratislava, Slovak Republic\\
$^{13}$ COMSATS University Islamabad, Islamabad, Pakistan\\
$^{14}$ Creighton University, Omaha, Nebraska, United States\\
$^{15}$ Department of Physics, Aligarh Muslim University, Aligarh, India\\
$^{16}$ Department of Physics, Pusan National University, Pusan, Republic of Korea\\
$^{17}$ Department of Physics, Sejong University, Seoul, Republic of Korea\\
$^{18}$ Department of Physics, University of California, Berkeley, California, United States\\
$^{19}$ Department of Physics, University of Oslo, Oslo, Norway\\
$^{20}$ Department of Physics and Technology, University of Bergen, Bergen, Norway\\
$^{21}$ Dipartimento di Fisica, Universit\`{a} di Pavia, Pavia, Italy\\
$^{22}$ Dipartimento di Fisica dell'Universit\`{a} and Sezione INFN, Cagliari, Italy\\
$^{23}$ Dipartimento di Fisica dell'Universit\`{a} and Sezione INFN, Trieste, Italy\\
$^{24}$ Dipartimento di Fisica dell'Universit\`{a} and Sezione INFN, Turin, Italy\\
$^{25}$ Dipartimento di Fisica e Astronomia dell'Universit\`{a} and Sezione INFN, Bologna, Italy\\
$^{26}$ Dipartimento di Fisica e Astronomia dell'Universit\`{a} and Sezione INFN, Catania, Italy\\
$^{27}$ Dipartimento di Fisica e Astronomia dell'Universit\`{a} and Sezione INFN, Padova, Italy\\
$^{28}$ Dipartimento di Fisica `E.R.~Caianiello' dell'Universit\`{a} and Gruppo Collegato INFN, Salerno, Italy\\
$^{29}$ Dipartimento DISAT del Politecnico and Sezione INFN, Turin, Italy\\
$^{30}$ Dipartimento di Scienze MIFT, Universit\`{a} di Messina, Messina, Italy\\
$^{31}$ Dipartimento Interateneo di Fisica `M.~Merlin' and Sezione INFN, Bari, Italy\\
$^{32}$ European Organization for Nuclear Research (CERN), Geneva, Switzerland\\
$^{33}$ Faculty of Electrical Engineering, Mechanical Engineering and Naval Architecture, University of Split, Split, Croatia\\
$^{34}$ Faculty of Engineering and Science, Western Norway University of Applied Sciences, Bergen, Norway\\
$^{35}$ Faculty of Nuclear Sciences and Physical Engineering, Czech Technical University in Prague, Prague, Czech Republic\\
$^{36}$ Faculty of Physics, Sofia University, Sofia, Bulgaria\\
$^{37}$ Faculty of Science, P.J.~\v{S}af\'{a}rik University, Ko\v{s}ice, Slovak Republic\\
$^{38}$ Frankfurt Institute for Advanced Studies, Johann Wolfgang Goethe-Universit\"{a}t Frankfurt, Frankfurt, Germany\\
$^{39}$ Fudan University, Shanghai, China\\
$^{40}$ Gangneung-Wonju National University, Gangneung, Republic of Korea\\
$^{41}$ Gauhati University, Department of Physics, Guwahati, India\\
$^{42}$ Helmholtz-Institut f\"{u}r Strahlen- und Kernphysik, Rheinische Friedrich-Wilhelms-Universit\"{a}t Bonn, Bonn, Germany\\
$^{43}$ Helsinki Institute of Physics (HIP), Helsinki, Finland\\
$^{44}$ High Energy Physics Group,  Universidad Aut\'{o}noma de Puebla, Puebla, Mexico\\
$^{45}$ Horia Hulubei National Institute of Physics and Nuclear Engineering, Bucharest, Romania\\
$^{46}$ Indian Institute of Technology Bombay (IIT), Mumbai, India\\
$^{47}$ Indian Institute of Technology Indore, Indore, India\\
$^{48}$ INFN, Laboratori Nazionali di Frascati, Frascati, Italy\\
$^{49}$ INFN, Sezione di Bari, Bari, Italy\\
$^{50}$ INFN, Sezione di Bologna, Bologna, Italy\\
$^{51}$ INFN, Sezione di Cagliari, Cagliari, Italy\\
$^{52}$ INFN, Sezione di Catania, Catania, Italy\\
$^{53}$ INFN, Sezione di Padova, Padova, Italy\\
$^{54}$ INFN, Sezione di Pavia, Pavia, Italy\\
$^{55}$ INFN, Sezione di Torino, Turin, Italy\\
$^{56}$ INFN, Sezione di Trieste, Trieste, Italy\\
$^{57}$ Inha University, Incheon, Republic of Korea\\
$^{58}$ Institute for Gravitational and Subatomic Physics (GRASP), Utrecht University/Nikhef, Utrecht, Netherlands\\
$^{59}$ Institute of Experimental Physics, Slovak Academy of Sciences, Ko\v{s}ice, Slovak Republic\\
$^{60}$ Institute of Physics, Homi Bhabha National Institute, Bhubaneswar, India\\
$^{61}$ Institute of Physics of the Czech Academy of Sciences, Prague, Czech Republic\\
$^{62}$ Institute of Space Science (ISS), Bucharest, Romania\\
$^{63}$ Institut f\"{u}r Kernphysik, Johann Wolfgang Goethe-Universit\"{a}t Frankfurt, Frankfurt, Germany\\
$^{64}$ Instituto de Ciencias Nucleares, Universidad Nacional Aut\'{o}noma de M\'{e}xico, Mexico City, Mexico\\
$^{65}$ Instituto de F\'{i}sica, Universidade Federal do Rio Grande do Sul (UFRGS), Porto Alegre, Brazil\\
$^{66}$ Instituto de F\'{\i}sica, Universidad Nacional Aut\'{o}noma de M\'{e}xico, Mexico City, Mexico\\
$^{67}$ iThemba LABS, National Research Foundation, Somerset West, South Africa\\
$^{68}$ Jeonbuk National University, Jeonju, Republic of Korea\\
$^{69}$ Johann-Wolfgang-Goethe Universit\"{a}t Frankfurt Institut f\"{u}r Informatik, Fachbereich Informatik und Mathematik, Frankfurt, Germany\\
$^{70}$ Korea Institute of Science and Technology Information, Daejeon, Republic of Korea\\
$^{71}$ KTO Karatay University, Konya, Turkey\\
$^{72}$ Laboratoire de Physique des 2 Infinis, Ir\`{e}ne Joliot-Curie, Orsay, France\\
$^{73}$ Laboratoire de Physique Subatomique et de Cosmologie, Universit\'{e} Grenoble-Alpes, CNRS-IN2P3, Grenoble, France\\
$^{74}$ Lawrence Berkeley National Laboratory, Berkeley, California, United States\\
$^{75}$ Lund University Department of Physics, Division of Particle Physics, Lund, Sweden\\
$^{76}$ Nagasaki Institute of Applied Science, Nagasaki, Japan\\
$^{77}$ Nara Women{'}s University (NWU), Nara, Japan\\
$^{78}$ National and Kapodistrian University of Athens, School of Science, Department of Physics , Athens, Greece\\
$^{79}$ National Centre for Nuclear Research, Warsaw, Poland\\
$^{80}$ National Institute of Science Education and Research, Homi Bhabha National Institute, Jatni, India\\
$^{81}$ National Nuclear Research Center, Baku, Azerbaijan\\
$^{82}$ National Research and Innovation Agency - BRIN, Jakarta, Indonesia\\
$^{83}$ Niels Bohr Institute, University of Copenhagen, Copenhagen, Denmark\\
$^{84}$ Nikhef, National institute for subatomic physics, Amsterdam, Netherlands\\
$^{85}$ Nuclear Physics Group, STFC Daresbury Laboratory, Daresbury, United Kingdom\\
$^{86}$ Nuclear Physics Institute of the Czech Academy of Sciences, Husinec-\v{R}e\v{z}, Czech Republic\\
$^{87}$ Oak Ridge National Laboratory, Oak Ridge, Tennessee, United States\\
$^{88}$ Ohio State University, Columbus, Ohio, United States\\
$^{89}$ Physics department, Faculty of science, University of Zagreb, Zagreb, Croatia\\
$^{90}$ Physics Department, Panjab University, Chandigarh, India\\
$^{91}$ Physics Department, University of Jammu, Jammu, India\\
$^{92}$ Physics Department, University of Rajasthan, Jaipur, India\\
$^{93}$ Physics Program and International Institute for Sustainability with Knotted Chiral Meta Matter (SKCM2), Hiroshima University, Hiroshima, Japan\\
$^{94}$ Physikalisches Institut, Eberhard-Karls-Universit\"{a}t T\"{u}bingen, T\"{u}bingen, Germany\\
$^{95}$ Physikalisches Institut, Ruprecht-Karls-Universit\"{a}t Heidelberg, Heidelberg, Germany\\
$^{96}$ Physik Department, Technische Universit\"{a}t M\"{u}nchen, Munich, Germany\\
$^{97}$ Politecnico di Bari and Sezione INFN, Bari, Italy\\
$^{98}$ Research Division and ExtreMe Matter Institute EMMI, GSI Helmholtzzentrum f\"ur Schwerionenforschung GmbH, Darmstadt, Germany\\
$^{99}$ Saha Institute of Nuclear Physics, Homi Bhabha National Institute, Kolkata, India\\
$^{100}$ School of Physics and Astronomy, University of Birmingham, Birmingham, United Kingdom\\
$^{101}$ Secci\'{o}n F\'{\i}sica, Departamento de Ciencias, Pontificia Universidad Cat\'{o}lica del Per\'{u}, Lima, Peru\\
$^{102}$ Stefan Meyer Institut f\"{u}r Subatomare Physik (SMI), Vienna, Austria\\
$^{103}$ SUBATECH, IMT Atlantique, Nantes Universit\'{e}, CNRS-IN2P3, Nantes, France\\
$^{104}$ Suranaree University of Technology, Nakhon Ratchasima, Thailand\\
$^{105}$ Technical University of Ko\v{s}ice, Ko\v{s}ice, Slovak Republic\\
$^{106}$ The Henryk Niewodniczanski Institute of Nuclear Physics, Polish Academy of Sciences, Cracow, Poland\\
$^{107}$ The University of Texas at Austin, Austin, Texas, United States\\
$^{108}$ Universidad Aut\'{o}noma de Sinaloa, Culiac\'{a}n, Mexico\\
$^{109}$ Universidade de S\~{a}o Paulo (USP), S\~{a}o Paulo, Brazil\\
$^{110}$ Universidade Estadual de Campinas (UNICAMP), Campinas, Brazil\\
$^{111}$ Universidade Federal do ABC, Santo Andre, Brazil\\
$^{112}$ University of Cape Town, Cape Town, South Africa\\
$^{113}$ University of Houston, Houston, Texas, United States\\
$^{114}$ University of Jyv\"{a}skyl\"{a}, Jyv\"{a}skyl\"{a}, Finland\\
$^{115}$ University of Kansas, Lawrence, Kansas, United States\\
$^{116}$ University of Liverpool, Liverpool, United Kingdom\\
$^{117}$ University of Science and Technology of China, Hefei, China\\
$^{118}$ University of South-Eastern Norway, Kongsberg, Norway\\
$^{119}$ University of Tennessee, Knoxville, Tennessee, United States\\
$^{120}$ University of the Witwatersrand, Johannesburg, South Africa\\
$^{121}$ University of Tokyo, Tokyo, Japan\\
$^{122}$ University of Tsukuba, Tsukuba, Japan\\
$^{123}$ University Politehnica of Bucharest, Bucharest, Romania\\
$^{124}$ Universit\'{e} Clermont Auvergne, CNRS/IN2P3, LPC, Clermont-Ferrand, France\\
$^{125}$ Universit\'{e} de Lyon, CNRS/IN2P3, Institut de Physique des 2 Infinis de Lyon, Lyon, France\\
$^{126}$ Universit\'{e} de Strasbourg, CNRS, IPHC UMR 7178, F-67000 Strasbourg, France, Strasbourg, France\\
$^{127}$ Universit\'{e} Paris-Saclay Centre d'Etudes de Saclay (CEA), IRFU, D\'{e}partment de Physique Nucl\'{e}aire (DPhN), Saclay, France\\
$^{128}$ Universit\`{a} degli Studi di Foggia, Foggia, Italy\\
$^{129}$ Universit\`{a} del Piemonte Orientale, Vercelli, Italy\\
$^{130}$ Universit\`{a} di Brescia, Brescia, Italy\\
$^{131}$ Variable Energy Cyclotron Centre, Homi Bhabha National Institute, Kolkata, India\\
$^{132}$ Warsaw University of Technology, Warsaw, Poland\\
$^{133}$ Wayne State University, Detroit, Michigan, United States\\
$^{134}$ Westf\"{a}lische Wilhelms-Universit\"{a}t M\"{u}nster, Institut f\"{u}r Kernphysik, M\"{u}nster, Germany\\
$^{135}$ Wigner Research Centre for Physics, Budapest, Hungary\\
$^{136}$ Yale University, New Haven, Connecticut, United States\\
$^{137}$ Yonsei University, Seoul, Republic of Korea\\
$^{138}$  Zentrum  f\"{u}r Technologie und Transfer (ZTT), Worms, Germany\\
$^{139}$ Affiliated with an institute covered by a cooperation agreement with CERN\\
$^{140}$ Affiliated with an international laboratory covered by a cooperation agreement with CERN.\\

\end{flushleft}

\end{document}